# Modeling Firm-Level ESG–Sentiment Interactions in Stock Returns: Evidence from 16 Companies Using Retrofitted Word Embeddings

Sangdeok Lee (kamgissang@unist.ac.kr)

**Abstract**

This study investigates how emotion-specific sentiment embedded in financial news headlines interacts with firm-level Environmental, Social, and Governance (ESG) ratings to influence stock return behavior. Addressing key methodological gaps in existing literature, the analysis leverages Retrofitted Word Embeddings to encode discrete emotional cues tailored to ESG-relevant narratives. Unlike prior studies that rely on lexicon-based or transformer-based models, this approach explicitly incorporates domain-specific emotional semantics while accounting for firm-level heterogeneity and temporal sentiment fluctuations.

Using a dataset of 16 multinational firms and sentiment data extracted from Seeking Alpha headlines, the study tests three hypotheses: (1) emotion-specific sentiment independently predicts stock returns; (2) the moderating effect of sentiment varies across ESG dimensions; and (3) positive (negative) sentiment amplifies (dampens) ESG performance effects. The analysis implements a dual sentiment aggregation strategy and introduces a triple-significance filtering criterion to identify robust interactions.

Results support Hypotheses 1 and 2, with emotions such as anticipation and trust showing consistent associations with return variation across firms and ESG categories. However, findings for Hypothesis 3 are mixed: while some sentiment–ESG combinations align with theoretical expectations, many contradictory interactions exhibit stronger effects. In this study, retrofitted embeddings outperform the NRC Emotion Lexicon in explaining stock return variation within ESG–sentiment interaction models, underscoring the value of emotional nuance in ESG-finance modeling. These results underscore the importance of emotion-sensitive sentiment modeling in understanding investor behavior and ESG-related stock price movements.

# Table of Context

# 1. Introduction

Environmental, Social, and Governance (ESG) considerations have become critical determinants of financial performance and investor decisions. However, the interpretation of ESG information by investors is not purely quantitative but is significantly influenced by the sentiment and emotional framing embedded in financial narratives, especially news headlines.

Traditional sentiment analysis methods—using lexicon-based models (e.g., NRC(National Research Council Canada) Emotion Lexicon [13]) or standard word embeddings (e.g., Word2Vec (Word to Vector) [10], GloVe (Global Vectors for Word Representation) [40], Doc2Vec (Document to Vector) [11])—lack the nuance required to capture emotion-specific sentiments that are particularly relevant to ESG contexts. Subtle emotional cues in ESG narratives shape investor behavior. However, current models lack the capacity to extract these emotion-specific signals in a firm- and time-reflective manner, posing a key methodological and empirical gap. Retrofitted Word Embeddings, which enhance conventional embeddings with emotional semantics from external lexicons, represent an unexplored yet promising avenue for addressing this gap [14]. Regardless, to the best of my knowledge, no existing study has empirically integrated emotion-specific Retrofitted Word Embeddings into ESG sentiment analysis to examine their predictive value on stock market outcomes. Furthermore, existing studies often analyze aggregate market sentiment or broad ESG indices, overlooking firm-level heterogeneity in sentiment effects. They also rarely address temporal bias introduced by uneven news volume, limiting the stability and interpretability of sentiment signals over time.

Previous research addressing sentiment analysis in ESG and financial contexts has predominantly relied on lexicon-based methods, such as the NRC Emotion Lexicon (Giatsoglou et al.) [12], or general-purpose word embeddings like Word2Vec (Deho et al.) [9]. These approaches primarily capture semantic meanings derived from large corpora but do not inherently reflect domain-specific emotional nuances crucial for ESG narratives. Some studies, notably Giatsoglou et al. (2017) [12], integrated general embeddings with emotion lexicons like NRC to better handle emotionally-rich texts, significantly enhancing sentiment prediction in financial news. More recently, researchers have shifted toward transformer-based language models such as BERT(Bidirectional Encoder Representations from Transformers) and FinBERT(Financial Language Representation BERT Model). For instance, Dorfleitner and Zhang (2024) [21] employed BERT to capture nuanced ESG-related sentiment signals, demonstrating predictive value for stock returns. Nevertheless, while transformer-based models offer context-sensitive sentiment analysis, they still depend on corpus-driven semantics without explicit emotional retrofitting. Consequently, despite the advances of transformer-based models, existing approaches have yet to explicitly leverage emotionally retrofitted embeddings that are both domain-specific and interpretable, especially in modeling how discrete emotions such as joy (happy), sadness (sad), anger (angry), fear, surprise, trust, disgust, and anticipation shape investor reactions to ESG disclosures at the firm level.

Although transformer-based models like BERT and general embeddings have significantly advanced sentiment analysis in ESG contexts, these methods do not explicitly incorporate emotional semantics tailored specifically to ESG-relevant narratives. The interpretation of ESG-related content by investors often hinges on subtle emotional signals—such as happy, sad, angry, fear, surprise, trust, disgust, and anticipation—that general-purpose embeddings may not sufficiently capture. Retrofitted Word Embeddings, by adjusting conventional embeddings to explicitly encode emotional relationships using domain-specific lexicons, offer an innovative and potentially more effective approach. By explicitly modeling emotional dimensions, Retrofitted Embeddings could provide deeper insights into how specific sentiments interact with ESG perceptions, thereby enhancing predictive power and interpretability in financial analyses. Therefore, exploring Retrofitted Word Embeddings represents a necessary advancement, addressing the emotional nuance gap left by existing NLP(Natural Language Processing) techniques in ESG-focused financial sentiment analysis.

This study aims to investigate whether integrating emotion-specific Retrofitted Word Embeddings into ESG-focused sentiment analysis enhances the predictive accuracy and explanatory power of stock return models. Specifically, the research examines how the emotional content of financial news headlines, captured through embeddings explicitly retrofitted for emotions, interacts with ESG scores to influence stock returns. By explicitly modeling emotional nuance in ESG communications, this study not only tests the predictive power of Retrofitted Word Embeddings but also conducts firm-level comparative modeling, a dual sentiment aggregation strategy, and triple-significance validation, offering a more robust framework to uncover how sentiment moderates the ESG–return relationship.

## 2. Theoretical and Empirical Background

### 2.1 Evolution of ESG Ratings

The use of ESG (Environmental, Social, and Governance) data in academic and financial contexts began in the late 20th century, driven by the rise of corporate social responsibility and the need for structured non-financial reporting. Early frameworks, such as the Global Reporting Initiative (GRI) launched in 1997, and data providers like Kinder, Lydenberg, and Domini (KLD) in 1988, laid the groundwork for systematic ESG assessment [1]. The 2004 "Who Cares Wins" report by the UN Global Compact formalized the ESG concept, advocating its integration into mainstream investment analysis [2]. Despite these advances, early ESG data was sparse and fragmented, and it suffered from standardization and transparency issues, challenges that persisted even as comprehensive datasets emerged from vendors like ASSET4 (later acquired by Thomson Reuters, now Refinitiv) [3].

### 2.2 ESG Ratings and Financial Market Behavior

Academic interest in the relationship between ESG ratings and stock prices has grown substantially. Initial studies were inconclusive due to inconsistent definitions and limited data, but recent meta-analyses (e.g., NYU Stern, 2021) show that a majority of studies find a positive correlation between ESG performance and financial returns [4][5]. Notably, the impact of ESG ratings on stock returns is moderated by the level of disagreement among rating providers, with greater disagreement being associated with higher perceived uncertainty and a corresponding return premium [6]. Furthermore, mediation analyses highlight that ESG's financial impact is often channeled through investor sentiment and disclosure practices, underscoring the complexity of these relationships [7].

### 2.3 The Rise of Sentiment Analysis

According to Google Trends data, interest in "sentiment analysis" has been increasing since 2004, gradually surpassing "customer feedback" in recent years (Google Trends, n.d.). While "customer feedback" initially dominated global search trends, the popularity of "sentiment analysis" began to rival and occasionally surpass it by approximately 2018. This rise coincides with the growing application of machine learning and AI in text analysis. Regionally, countries such as Iran, Tunisia, and China show a higher relative interest in sentiment analysis, while Western nations like the United States and Australia maintain a stronger interest in traditional customer feedback. These patterns suggest a global shift toward computational approaches in interpreting consumer sentiment [8].

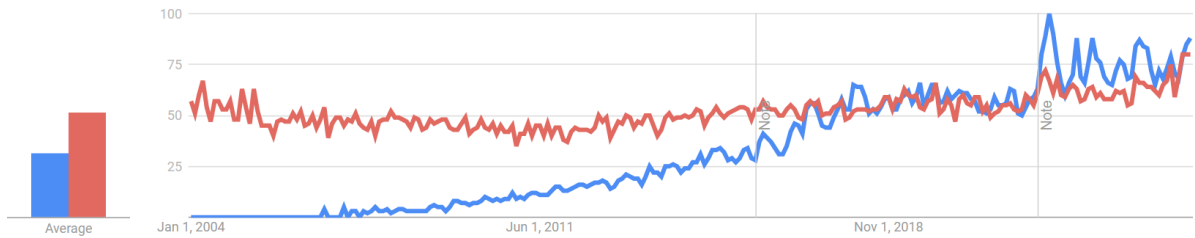

**Figure 2.3.a Google Trends Comparison of Search Interest in "ESG" (blue) and "CSR" (red), 2004–2024**

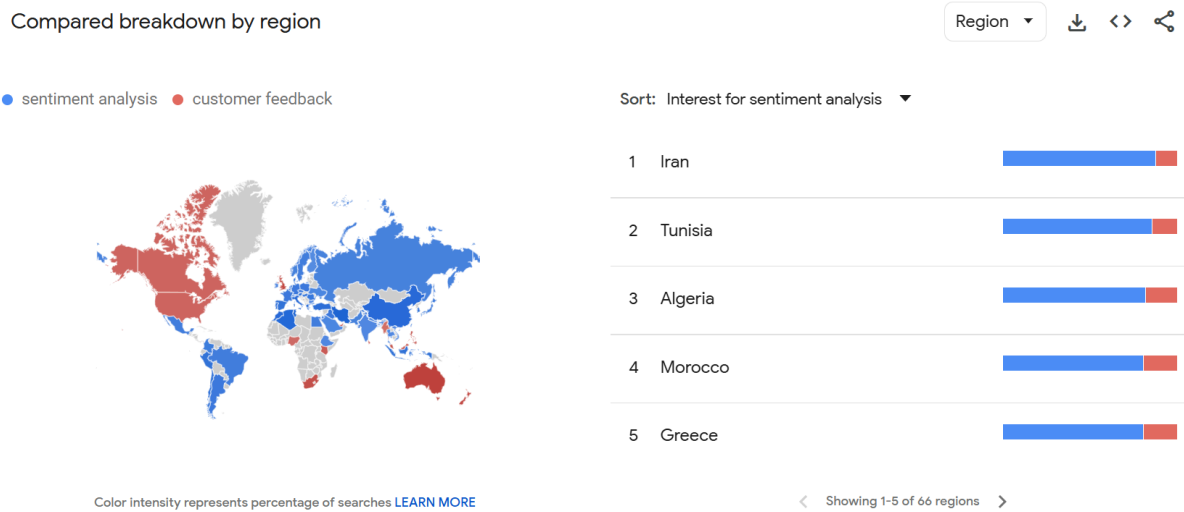

**Figure 2.3.b Global map and top regional rankings by search interest for "sentiment analysis" versus "customer feedback." Color intensity indicates relative search volume.**

### 2.3.1 Historical Development and Technical Advancements in Sentiment Analysis

Sentiment analysis has evolved from rule-based approaches and static lexicons to more sophisticated methods powered by distributed word representations. Early models, while interpretable, lacked scalability and contextual understanding. The introduction of Word2Vec and Doc2Vec marked a turning point by enabling words and documents to be mapped into continuous vector spaces that capture semantic and syntactic relationships (Deho et al., 2018; Mikolov et al., 2013; Le et al., 2014) [9][10][11].

To enhance emotional sensitivity, researchers have integrated affective resources into embedding models. Giatsoglou et al. (2017) [12] combined Word2Vec with the NRC Emotion Lexicon to capture specific emotions like anger, joy, and sadness, improving performance on emotionally rich texts such as news headlines. Shah et al. (2022) [15][16] further advanced this by retrofitting pre-trained embeddings with external emotional knowledge, yielding better sentiment detection in both short and long texts. These innovations have made sentiment analysis more context-aware and adaptable, especially in domains like finance and ESG, where nuanced emotional signals carry interpretive weight.

## 2.3.2 From Lexicon-Based Sentiment to Emotion-Enriched Embeddings

Early sentiment analysis in finance and social sciences frequently relied on lexicon-based approaches, with the NRC Emotion Lexicon (Mohammad & Turney, 2013) [13] being among the most widely adopted tools. These lexicons offer a straightforward mechanism for emotion detection by associating individual words with discrete emotional categories (e.g., trust, fear, joy, disgust). However, their binary labeling structure lacks sensitivity to contextual nuance and semantic proximity, which are essential for capturing complex emotional tones in financial narratives.

To address these limitations, more recent studies have turned to vector-based representations that incorporate emotional semantics into distributed word embeddings. A pivotal advancement in this direction was the retrofitting method proposed by Faruqui et al. (2015) [14], which adjusts pre-trained embeddings (e.g., GloVe, Word2Vec) using external lexical knowledge, including synonym sets and semantic constraints. This technique has since been adapted for sentiment analysis through emotion-specific retrofitting, where embeddings are fine-tuned using emotion lexicons like NRC to produce dense vector representations aligned with distinct emotional states. Although retrofitting was first introduced in 2015 [14], its application to emotion-enriched sentiment modeling gained significant traction in the early 2020s, particularly in finance contexts where emotional subtleties in news content have measurable effects on market behavior (Shah, Reddy, and Bhattacharyya (2022a, 2022b) [15][16]. This transition reflects a broader methodological evolution—from rigid, dictionary-based classification to flexible, semantically informed sentiment modeling.

**Table2.3.2. Comparison of NRC Emotion Lexicon and Retrofitted Emotion Embeddings**

| Aspect | NRC Emotion Lexicon | Retrofitted Emotion Embeddings |
|---|---|---|
| **Methodological Basis** | Dictionary-based, manual annotations | Embedding-based, lexical retrofitting |
| **Emotional Representation** | Binary presence or absence of emotions | Continuous, vector-based emotional semantics |
| **Semantic Richness** | Limited (categorical) | High (nuanced semantic relations) |
| **Interpretability** | High | Moderate (embedding complexity) |
| **Implementation Complexity** | Low (simple lookup) | Moderate (requires embedding optimization) |
| **Flexibility & Customization** | Limited (fixed categories) | Highly customizable (adaptable embeddings) |
| **Suitable Use Cases** | Quick emotional categorization tasks, baseline sentiment analysis | Advanced sentiment analysis tasks, nuanced contexts |
| **Computational Requirement** | Low | Moderate to high |

Table 2.3.2 summarizes the conceptual and methodological differences between the NRC Emotion Lexicon and Retrofitted Emotion Embeddings. While both approaches aim to identify emotional tone in text, they diverge in structure and analytical capability. The NRC Lexicon relies on a fixed, binary association between words and emotion categories, offering interpretability but limited contextual sensitivity. In contrast, Retrofitted Emotion Embeddings encode emotional semantics into continuous vector spaces by adjusting pre-trained word embeddings using emotion-based lexical constraints. This allows for more nuanced sentiment quantification, particularly in domains like financial news where semantic subtlety matters. The table highlights key trade-offs in interpretability, granularity, flexibility, and suitability for advanced sentiment modeling, justifying the use of retrofitted embeddings in this study's ESG-related textual analysis.

### 2.3.3 Retrofitted Word Embeddings in Financial and ESG Research

Retrofitted Word Embeddings represent a significant advancement in sentiment analysis, particularly within financial and ESG-related contexts. Unlike traditional distributional models such as Word2Vec or GloVe, which derive word meanings solely from co-occurrence patterns in large corpora (Mikolov et al., 2013) [17][40], retrofitted embeddings incorporate additional semantic or affective constraints from external lexicons. This post hoc adjustment allows the vectors to reflect better domain-specific nuances such as emotion, tone, or investor sentiment.

The retrofitting technique, introduced by Faruqui et al. (2015) [14], modifies pre-trained word vectors by aligning them with semantic lexicons like WordNet or PPDB (Paraphrase Database) through a graph-based belief propagation process. This method ensures that words with known semantic relationships have more similar vector representations, improving performance on a variety of linguistic and sentiment-related tasks.

The original algorithm was introduced by Faruqui et al. (2015), and it aims to minimize the following objective:

$$\Psi = \sum_{i=1}^{|V|} \left[ \alpha_i \|\hat{q}_i - q_i\|^2 + \sum_{(i,j) \in E} \beta_{ij} \|\hat{q}_i - \hat{q}_j\|^2 \right]$$

Where:

- $q_i$ = original embedding of word *i*
- $\hat{q}_i$ = retrofitted embedding (to be learned)
- $E$ = set of word pairs with known relationships (e.g., synonyms)
- The first term ensures retrofitted vectors don't deviate too far from the original embeddings.
- The second term encourages semantically similar words to be close together.

This approach has been further adapted for effective modeling in financial contexts. Shah, Reddy, and Bhattacharyya (2022a, 2022b) [15][16] demonstrate that emotion-enriched retrofitted embeddings significantly outperform traditional embeddings in downstream applications such as financial sentiment detection, sarcasm recognition, and emotion classification. These enhancements make retrofitted embeddings particularly valuable in ESG research, where investor behavior is often influenced by subtle emotional cues embedded in textual disclosures.

By reshaping pre-trained vectors to encode affective content, retrofitted embeddings provide a more accurate representation of sentiment signals in ESG-related texts. This is essential for modeling how stakeholders interpret sustainability communications, offering a high-resolution lens into market sentiment and valuation dynamics.

## 2.4 Sentiment Analysis in News Media

Early sentiment analysis of news media has become an essential tool for understanding short-term stock price movements. Tetlock (2007) [18] was among the first to show that pessimistic language in *The Wall Street Journal* could predict short-term market declines and increased trading volume, findings that align with behavioral theories such as those proposed by DeLong et al. (1990) [19]. Since then, sentiment analysis has evolved into a core natural language processing (NLP) technique for extracting subjective information from text. In the context of news, it is used to uncover biases, gauge tone, and assess the emotional impact of reporting—factors that can influence public opinion and market behavior (Mäntylä et al., 2016) [20].

Recent advances in NLP have introduced transformer-based models like BERT and FinBERT, which are capable of capturing more nuanced sentiment in financial news. For example, Dorfleitner and Zhang (2024) [21] applied BERT to ESG-related news and found that positive sentiment was associated with abnormal returns of +0.31%, while negative sentiment led to −0.75%, with firm-level ESG scores moderating these effects. Similarly, Lee, Kim, and Park (2024) [22] applied FinBERT to analyze ESG-related sentiment from LexisNexis news articles and integrated the resulting sentiment index with technical indicators to enhance S&P 500 index prediction. Their deep learning model, specifically a Bi-LSTM architecture, demonstrated significantly improved forecasting accuracy—achieving a MAPE of 3.05%—when combining sentiment, price, and technical features, compared to models using only price data or technical indicators.

## 2.5 Media Framing and Investor Behavior

The influence of financial media on investor behavior is not limited to the factual content of articles but extends to how information is framed. According to Entman's (1993) [23] framing theory, media outlets emphasize certain aspects of reality while omitting others, thereby shaping audience interpretation through subtle cues such as emotional tone, word choice, and narrative structure. In the financial context, this framing is often reflected in headlines, which serve as cognitive shortcuts for readers, especially in high-frequency or algorithm-driven trading environments (Biais et al., 2011; Linstrõm et al., 2012; Beschwitz et al., 2015) [24][25][26].

Behavioral finance research further supports this mechanism. Investors frequently rely on heuristic-driven processes when interpreting complex or uncertain information (Barberis et al., 1998; Kahneman & Tversky, 1974) [27][28]. Emotional framing — such as expressing fear, optimism, or uncertainty — can influence perceived risk and reward, thus impacting investment decisions even in the absence of fundamental valuation changes. These insights help explain why emotionally charged headlines, as captured through retrofitted word embeddings and sentiment scores, may correlate with market movements and ESG perceptions, reinforcing the importance of studying emotion-laden narratives in financial media.

## 2.6 Seeking Alpha as a Behavioral Data Source in Finance

User-generated platforms like Seeking Alpha have become valuable tools in behavioral finance, offering timely, crowd-sourced investment analysis that reflects investor sentiment and market expectations. Its hybrid contributor base—comprising both professionals and amateurs—produces semi-structured content increasingly used in empirical research.

Pei, Anand, and Huan (2024) [29] show that sentiment from Seeking Alpha articles is significantly associated with short-term and drift-adjusted stock returns, outperforming traditional financial media in predictive power. Their results also indicate that sentiment affects option market dynamics, as seen in changes in implied volatility and skew.

Breuer, Knetsch, and Sachsenhausen (2024) [30] find that the readability and presentation of Seeking Alpha articles significantly influence investor behavior, with more accessible articles—those easier to read and less technical—linked to increased trading volume and short-term abnormal returns. These findings align with complexity aversion theory, where investors favor information that reduces cognitive load.

Supporting this, Lachana and Schröder (2024) [31] report that Seeking Alpha sentiment better explains stock return variance than sentiment from legacy news outlets, highlighting the platform's relevance for financial modeling.

Collectively, these studies establish Seeking Alpha as a credible behavioral data source, offering unique value for sentiment-aware financial and ESG research.

## 2.7 Integrating ESG Metrics and Textual Sentiment

ESG ratings offer structured evaluations of firms' sustainability performance, but they often vary significantly across different providers due to methodological differences. Integrating sentiment analysis with ESG ratings provides a complementary perspective by capturing real-time market perceptions.

Kvam et al. (2024) [32] showed that ESG sentiment derived from sources like Google and Twitter influences short-term stock returns, especially during periods of heightened ESG attention. During such times, firms with high ESG ratings tend to experience stronger positive market reactions, reflecting investors' preference for sustainability when ESG issues are in the spotlight. Likewise, Dorfleitner and Zhang (2024) [21] found that companies with robust ESG profiles are less affected by negative news and benefit more from positive sentiment, indicating that ESG scores can buffer or amplify the impact of news depending on its tone.

This integration allows for more dynamic modeling of ESG impacts by considering both formal sustainability metrics and real-time investor sentiment. As demonstrated by Lee et al. (2024) [22][33], combining ESG indicators with NLP-driven sentiment analysis leads to improved predictive performance in stock forecasting and automated ESG grading. These findings underscore the value of hybrid approaches in advancing ESG finance research.

## 2.8 Hypotheses Development

While ESG performance has traditionally served as a static indicator of firm-level sustainability, recent developments in behavioral finance and natural language processing emphasize the importance of how such performance is perceived by the public. Sentiment embedded in financial news—especially in headlines—acts as a behavioral filter, shaping investor interpretations and reactions to ESG signals. This study integrates emotion-specific sentiment metrics derived from retrofitted word embeddings with ESG ratings to examine their combined influence on stock return behavior. Prior research has shown that investor sentiment captured through financial news can predict return volatility and market anomalies (Tetlock, 2007) [18], and that firm-level ESG perception plays a growing role in shaping investor decisions (Serafeim & Yoon, 2021) [34].

### 2.8.1 H1: The Influence of Emotion-Specific Sentiment on Stock Returns

Investor behavior is increasingly influenced by emotionally charged narratives in financial media. Sentiments such as *Happy*, *Sad*, *Angry*, *Fear*, *Surprise*, *Trust*, *Disgust*, and *Anticipation*, when embedded in headline content, are believed to shape market expectations and reactions, especially during periods of uncertainty. These emotion-specific signals provide additional explanatory power beyond traditional financial indicators and basic sentiment polarity. Prior studies have demonstrated that financial news tone and emotionality significantly impact short-term asset returns (Heston & Sinha, 2016) [35].

To capture this effect, this study includes interaction terms between ESG scores and sentiment in the regression framework:

$$ESASR_{i,t} = \alpha + \beta_1 \cdot ESG_{i,t} + \beta_2 \cdot Sentiment^{InpageTitle}_{i,t} + \beta_3 \cdot \left( ESG_{i,t} \times Sentiment^{InpageTitle}_{i,t} \right) + \varepsilon_{i,t}$$

Where:

- $ESASR_{i,t}$ Stock return for firm *i* at time *t*, adjusted by ESG and sentiment interaction.
- $ESG_{i,t}$ : ESG score of firm *i* at time *t*.
- $Sentiment^{InpageTitle}_{i,t}$ : Sentiment is the cosine similarity score between news headlines("InpageTitle") and a retrofitted emotion vector.
- $ESG_{i,t} \times Sentiment^{InpageTitle}_{i,t}$ : Captures moderation—i.e., whether sentiment amplifies or dampens the market effect of ESG.

**Hypothesis 1.** Emotion-specific sentiment embedded in financial news headlines is significantly associated with future stock returns.

### 2.8.2 H2: Variation Across ESG Dimensions

Environmental, social, and governance indicators may exhibit differing levels of sensitivity to sentiment cues. For instance, *social* sentiment may carry greater relevance in consumer-facing sectors, while *governance* sentiment may be more critical in highly regulated industries such as finance or healthcare. Empirical studies support the idea that ESG subdimensions show heterogeneous associations with financial outcomes—Serafeim and Yoon (2021) [34] report that social capital and product quality news elicit stronger investor responses than environmental news, while Cek and Eyupoglu (2020) [36] find that governance performance significantly affects economic performance.

**Hypothesis 2.** The moderating effect of sentiment on ESG-related stock returns varies across ESG subdimensions (Environmental, Social, and Governance).

### 2.8.3 H3: Sentiment as a Moderator of ESG–Return Relationships

The impact of ESG performance on financial outcomes is not merely a function of the scores themselves, but also of how those scores are contextualized in public discourse. For instance, negative sentiment (e.g., *disgust*) may diminish the market benefits of high governance performance, while positive sentiment (e.g., *trust*) may amplify the value attributed to strong social practices. This perspective aligns with recent findings that sentiment derived from ESG-related news can significantly influence how markets interpret ESG disclosures (Dorfleitner & Zhang, 2024; Yu et al., 2023) [21][37].

**Hypothesis 3.** The effect of ESG scores on stock returns is moderated by the sentiment of financial news headlines; positive (negative) sentiment enhances (diminishes) the impact of ESG performance.

## 3. Methodology

### 3.1 Overview of Analytical Framework

This study investigates the relationship between emotional tone in financial news and firm-level ESG indicators by leveraging retrofitted word embeddings and statistical correlation analysis. The central premise is that emotionally charged language in financial texts—particularly in titles and headlines—can reflect underlying investor sentiment, which may be associated with ESG performance and financial outcomes.

To explore this, the study applies emotion-specific retrofitted word embeddings to news data sourced from Seeking Alpha and aligns these features with firm-level ESG metrics. This streamlined framework enables a focused analysis of how latent emotional signals embedded in financial language relate to corporate sustainability and market behavior.

### 3.2 Data Collection and Preprocessing

This study constructs a comprehensive textual dataset comprising financial articles and headlines sourced from Seeking Alpha, covering the period from 2008 to 2020. The dataset focuses on 16 multinational firms, selected based on the availability of consistent ESG score data from Refinitiv. These firms span a range of industries—including technology, energy, healthcare, and finance—to ensure sectoral diversity and allow for the examination of cross-industry generalizability in sentiment effects.

Data were collected using a custom web crawler developed in Python, utilizing Selenium and BeautifulSoup to systematically extract article headlines and associated metadata. The crawler was designed to exclude duplicates and non-financial content to maintain topical relevance. All textual data were subjected to a standardized preprocessing pipeline consisting of tokenization, lowercasing, stop-word removal, and lemmatization. This process ensured uniform text formatting, thereby enhancing the accuracy of downstream emotional similarity scoring using retrofitted word embeddings.

### Table 3.2.a. Company Ticker and Full Name Mapping

| Ticker (Refinitiv) | Ticker (SeekingAlpha/Yahoo Finance) | Company Name |
|---|---|---|
| AAPL | AAPL | Apple Inc. |
| AMZN | AMZN | Amazon.com, Inc. |
| BA | BA | The Boeing Company |
| BAC | BAC | Bank of America Corporation |
| BPORD | BP | BP p.l.c. (Ordinary Shares) |
| BNPQ | BNPQY | BNP Paribas SA |
| CVS | CVS | CVS Health Corporation |
| F | F | Ford Motor Company |
| GS | GS | The Goldman Sachs Group, Inc. |
| TOYOT | TM | Toyota Motor Corporation |
| TENCE | TCEHY | Tencent Holdings Limited |
| NOKIA | NOK | Nokia Corporation |
| PFE | PFE | Pfizer Inc. |
| SALLI | BABA | Alibaba Pictures Group Ltd |
| SAMSU | SSNLF | Samsung Electronics Co., Ltd. |
| TSLA | TSLA | Tesla, Inc. |
| AllFirms (not from Refinitiv) | | Combined Market Dataset (Synthetic Aggregate) |

### Table 3.2.b. Annual Distribution of Crawled Articles

| Year | AAPL | AMZN | BA | BAC | BNPQ | BPORD | CVS | F | GS | NOKIA | PFE | SALLI | SAMSU | TENCE | TOYOT | TSLA | Total |
|---|---|---|---|---|---|---|---|---|---|---|---|---|---|---|---|---|---|
| 2008 | 15 | 1 | 94 | 156 | 4 | 36 | 11 | 96 | 137 | 74 | 49 | 0 | 0 | 0 | 58 | 0 | 731 |
| 2009 | 92 | 47 | 196 | 609 | 3 | 37 | 21 | 137 | 291 | 102 | 158 | 0 | 0 | 0 | 71 | 0 | 1764 |
| 2010 | 345 | 71 | 147 | 366 | 5 | 406 | 20 | 123 | 402 | 211 | 123 | 0 | 131 | 8 | 189 | 11 | 2558 |
| 2011 | 697 | 207 | 217 | 656 | 96 | 241 | 35 | 183 | 347 | 426 | 130 | 0 | 234 | 18 | 131 | 41 | 3659 |
| 2012 | 1072 | 317 | 222 | 341 | 23 | 285 | 48 | 297 | 240 | 471 | 222 | 0 | 294 | 25 | 182 | 142 | 4181 |
| 2013 | 642 | 291 | 381 | 287 | 7 | 324 | 48 | 338 | 228 | 288 | 193 | 0 | 202 | 35 | 279 | 396 | 3939 |
| 2014 | 229 | 201 | 221 | 189 | 56 | 218 | 43 | 268 | 164 | 67 | 147 | 122 | 92 | 41 | 207 | 298 | 2563 |
| 2015 | 181 | 162 | 260 | 146 | 12 | 246 | 40 | 206 | 116 | 65 | 186 | 108 | 75 | 43 | 161 | 277 | 2284 |
| 2016 | 309 | 262 | 299 | 165 | 9 | 188 | 31 | 210 | 161 | 47 | 166 | 79 | 117 | 41 | 141 | 380 | 2605 |
| 2017 | 523 | 489 | 333 | 153 | 16 | 161 | 57 | 233 | 142 | 52 | 189 | 142 | 194 | 108 | 142 | 354 | 3288 |
| 2018 | 521 | 591 | 390 | 138 | 16 | 175 | 79 | 239 | 192 | 63 | 236 | 158 | 134 | 161 | 130 | 504 | 3727 |
| 2019 | 417 | 335 | 492 | 157 | 20 | 183 | 92 | 207 | 209 | 88 | 213 | 111 | 101 | 110 | 101 | 398 | 3234 |
| 2020 | 473 | 555 | 457 | 205 | 22 | 204 | 78 | 229 | 225 | 110 | 417 | 167 | 81 | 143 | 107 | 563 | 4036 |
| Total | 5516 | 3529 | 3709 | 3568 | 289 | 2704 | 603 | 2766 | 2854 | 2064 | 2429 | 887 | 1655 | 733 | 1899 | 3364 | 38569 |

Table 3.2.b presents the yearly distribution of articles collected for analysis, spanning from 2008 to 2020. A total of 38,569 observations were recorded. Some zero entries in the annual article distribution table are attributable to the timeline of company events. For instance, Alibaba's IPO on September 19, 2014, implies that any articles referencing it in a corporate context would only begin to appear meaningfully from 2014 onward [38]. Similarly, Tesla IPOed on June 29, 2010, had limited visibility in earlier years due to its emerging status and smaller media footprint [39]. Thus, the absence of articles in certain early years reflects the natural lag in public and media engagement following foundational or IPO events. Nonetheless, it is important to note that not all instances of zero observations can be directly explained by company-specific milestones or known historical events and may also stem from limitations in data coverage, source availability, or indexing practices.

The final set of firms includes: AAPL, AMZN, BA, BAC, BPORD, BNPQ, CVS, F, GS, TOYOT, TENCE, NOKIA, PFE, SALLI, SAMSU, and TSLA. This selection enables robust control for industry-specific effects and facilitates a nuanced exploration of whether textual sentiment meaningfully interacts with ESG performance across heterogeneous disclosure environments.

To complement firm-level analysis and capture broader market patterns, an additional synthetic ticker labeled "AllFirms" was constructed. This aggregate represents a pooled regression combining all firm data to evaluate generalizable sentiment–ESG interaction effects across the entire dataset.

**Table 3.2.c. ESG and Sentiment Variables**

| Type | Column Name | Description |
|---|---|---|
| Text | InpageTitle_Retro_happy_similarity | Cosine similarity between yearly collective title vector and "happy" vector via Retrofitted Word Embeddings |
| Text | InpageTitle_Retro_sad_similarity | Cosine similarity between yearly collective title vector and "sad" vector via Retrofitted Word Embeddings |
| Text | InpageTitle_Retro_angry_similarity | Cosine similarity between yearly collective title vector and "angry" vector via Retrofitted Word Embeddings |
| Text | InpageTitle_Retro_fear_similarity | Cosine similarity between yearly collective title vector and "fear" vector via Retrofitted Word Embeddings |
| Text | InpageTitle_Retro_surprise_similarity | Cosine similarity between yearly collective title vector and "surprise" vector via Retrofitted Word Embeddings |
| Text | InpageTitle_Retro_trust_similarity | Cosine similarity between yearly collective title vector and "trust" vector via Retrofitted Word Embeddings |
| Text | InpageTitle_Retro_disgust_similarity | Cosine similarity between yearly collective title vector and "disgust" vector via Retrofitted Word Embeddings |
| Text | InpageTitle_Retro_anticipation_similarity | Cosine similarity between yearly collective title vector and "anticipation" vector via Retrofitted Word Embeddings |
| Text | InpageTitle_Retro_happy_similarity_DAvg | Daily average of cosine similarity between title vector and "happy" vector via retrofitted word embeddings |
| Text | InpageTitle_Retro_sad_similarity_DAvg | Daily average of cosine similarity between title vector and "sad" vector via retrofitted word embeddings |
| Text | InpageTitle_Retro_angry_similarity_DAvg | Daily average of cosine similarity between title vector and "angry" vector via retrofitted word embeddings |
| Text | InpageTitle_Retro_fear_similarity_DAvg | Daily average of cosine similarity between title vector and "fear" vector via retrofitted word embeddings |
| Text | InpageTitle_Retro_surprise_similarity_DAvg | Daily average of cosine similarity between title vector and "surprise" vector via retrofitted word embeddings |
| Text | InpageTitle_Retro_trust_similarity_DAvg | Daily average of cosine similarity between title vector and "trust" vector via retrofitted word embeddings |
| Text | InpageTitle_Retro_disgust_similarity_DAvg | Daily average of cosine similarity between title vector and "disgust" vector via retrofitted word embeddings |
| Text | InpageTitle_Retro_anticipation_similarity_DAvg | Daily average of cosine similarity between title vector and "anticipation" vector via retrofitted word embeddings |
| ESG | overall_score | Overall Score (overall_score) |
| ESG | econ_score | Economic Score (econ_score) |
| ESG | envrn_score | Environmental Score (envrn_score) |
| ESG | corpgov_score | Corporate Governance Score (corpgov_score) |
| ESG | social_score | Social Score (social_score) |
| Stock | Adj Close Adjusted close price adjusted for splits and dividend and/or capital gain distributions._R_DAvg | Daily average of the increase of the closing price adjusted for stock splits, dividends, and/or capital gain distributions compared to previous market open day |

Table 3.2.c summarizes the key variables used in the analysis, including ESG scores and sentiment similarity metrics derived from retrofitted embeddings.

In contrast to Dorfleitner and Zhang (2024) [21], who removed articles with cosine similarity scores exceeding 0.8 to reduce redundancy, this study retained such entries. This decision is grounded in behavioral finance theory, which suggests that repeated exposure to a particular narrative—especially through platforms like Seeking Alpha—may reinforce the perceived significance of the event. Rather than treating repetition as noise, this study interprets multiple similar articles as indicators of heightened investor attention or narrative salience, both of which can meaningfully influence market responses.

## 3.3 Construction of Emotion-Enriched Embeddings via Retrofitting

To model the emotional tone embedded in financial narratives, this study constructed sentiment-aware word representations using retrofitted word embeddings, an approach that modifies pre-trained vectors to better encode affective meaning. This method enhances the semantic relevance of embeddings, allowing for a more precise measurement of emotional content in financial texts.

The process began with loading 300-dimensional GloVe embeddings, originally trained on six billion tokens from Wikipedia and Gigaword, provided by the Stanford NLP Group [40]. These embeddings were then retrofitted using manually curated synonym sets corresponding to eight core emotions—joy (happy), sadness (sad), anger (angry), fear, surprise, trust, disgust, and anticipation. For each emotion, a list of closely related synonyms (e.g., for 'sad': 'unhappy', 'sorrowful', 'depressed') was created to serve as semantic constraints in the retrofitting process.

**Table 3.3.a Emotion Categories and Associated Synonyms Used for Retrofitting**

| Emotion | Synonyms |
|---|---|
| Happy | joyful, content, cheerful |
| Sad | unhappy, sorrowful, depressed |
| Angry | mad, furious, irritated |
| Fear | scared, afraid, terrified |
| Surprise | amazed, shocked, astonished |
| Trust | confident, hopeful, assured |
| Disgust | revolted, repelled, nauseated |
| Anticipation | expectation, hope, looking forward |

Following the method proposed by Faruqui et al. (2015) [14], retrofitting was conducted iteratively over 10 cycles. In each iteration, the vector for a target emotion word was updated as the mean of its current vector and the vectors of its defined semantic neighbors. This study implemented this procedure using Gensim's 'KeyedVectors' structure to ensure compatibility and memory efficiency. The final emotion-adjusted embeddings served as the foundation for emotional similarity scoring.

To quantify the alignment between news headlines and emotional tone, cosine similarity scores were calculated between the average embedding vector of each headline and each of the eight retrofitted emotion vectors. This yielded emotion-specific metrics such as 'InpageTitle_Retro_sad_similarity', capturing the affective tone in the text. These scores were computed for every document in the dataset and aggregated at the firm-month level for use in regression analysis.

This retrofitting technique provided a distinct advantage over static embeddings by reshaping vector spaces to reflect affective structure, thereby enabling more interpretable and targeted sentiment modeling in the context of ESG–financial performance analysis.

To account for temporal variations and ensure more robust sentiment measures, these emotion similarity scores were also aggregated into daily averages before being averaged again across the year for each firm, resulting in the "_Davg" variables (e.g., InpageTitle_Retro_sad_similarity_DAvg). This two-level aggregation was intended to minimize the influence of news volume fluctuations and provide a more stable estimate of the emotional tone experienced over time.

**Table 3.3.b Comparison of Sentiment Calculation Methods**

| Aspect | Non-_Davg Sentiment | _Davg Sentiment |
|---|---|---|
| Equation | $Similarity_{i,k} = \frac{1}{N_i} \sum_{j=1}^{N_i} \cos(H_{i,j}, E_k)$<br><br>$H_{i,j}$: embedding vector for the $j$-th headline of firm $i$<br><br>$E_k$: retrofitted embedding for emotion $k$<br><br>$\cos(\cdot,\cdot)$: cosine similarity<br><br>$N_i$: total number of headlines for firm $i$ | $Similarity\_DAvg_{i,k} = \frac{1}{T_i} \sum_{t=1}^{T_i} \frac{1}{|D_{i,t}|} \sum_{j \in D_{i,t}} \cos(H_{i,j}, E_k)$<br><br>$H_{i,j}$: embedding vector for the $j$-th headline of firm $i$<br><br>$E_k$: retrofitted embedding for emotion $k$<br><br>$\cos(\cdot,\cdot)$: cosine similarity<br><br>$D_{i,t}$: the set of headlines for firm $i$ on day $t$.<br><br>$T_i$: total number of days with headlines for firm $i$ |
| Headline Unit | Individual headline | Daily group of headlines |
| Aggregation Level | All headlines over the year | First by day, then by year |
| Temporal Normalization | No — biased by news volume | Yes — equal weight per day |
| Sensitivity to News Spikes | High — sentiment skewed if a few days have many headlines | Low — each day contributes equally |
| Use Case | Captures total emotional exposure of the year | Captures average daily emotional tone of the year |
| Interpretability | May reflect media intensity | Better reflects consistent sentiment trends over time |
| Best For | High-signal, event-driven studies | Long-term trend analysis, hypothesis testing |

## 3.4 ESG Variable Selection and Imputation Techniques

Firm-level ESG data were obtained from Refinitiv, including both composite ESG scores and disaggregated sub-scores across environmental, social, governance, and economic dimensions. Due to inconsistencies in ESG disclosure across firms and years, the dataset exhibited non-trivial levels of missingness, necessitating rigorous imputation prior to analysis.

To address this, this study applied Multiple Imputation by Chained Equations (MICE) using Predictive Mean Matching (PMM) [41], a method shown to preserve both the distributional properties and inter-variable relationships of heterogeneous datasets.

Imputations were performed separately for each firm-year panel to maintain both temporal and firm-specific integrity. Only variables with ≤50% missingness were retained, consistent with findings from longitudinal health research indicating that MICE yields valid inferences below this threshold (Pop Health Metrics, 2025) [42]. Variables exceeding this threshold were excluded due to the increased risk of bias and imprecision in imputed estimates.

The decision to use MICE with PMM was guided by prior ESG-focused imputation studies (Caprioli, 2024; Kotsantonis, 2019) [43][44], which emphasize its advantages in preserving realistic data structures within complex financial environments. Alternative methods—such as k-Nearest Neighbors (KNN), Random Forest, and regression-based imputation—were evaluated but ultimately rejected due to concerns over overfitting, computational cost, and variance distortion. Likewise, simpler techniques like mean or median imputation were excluded for their inability to maintain multivariate dependencies, a critical requirement in ESG modeling.

**Table 3.4 Comparative Overview of Missing Data Imputation Methods**

| Method | Pros | Cons |
|---|---|---|
| MICE with PMM | Preserves data distribution; accounts for uncertainty | Computationally intensive; assumes MAR (Missing At Random) |
| KNN Imputation | Simple; non-parametric | Sensitive to 'k' choice; less effective in high dimensions |
| Random Forest Imputation | Captures complex relationships; robust | Computationally demanding; may not suit small datasets |
| Mean/Median Imputation | Easy to implement; fast | Ignores data variability; can bias estimates |
| Regression Imputation | Utilizes variable relationships | Assumes linearity; doesn't account for prediction uncertainty |
| Expectation-Maximization (EM) | Provides maximum likelihood estimates | Assumes data distribution; risk of converging to local maxima |
| Deep Learning Imputation | Handles complex, high-dimensional data | Requires significant resources; less interpretable |

## 3.5 Stock Price and Exchange Rate Data Collection

To complement the ESG and sentiment datasets, historical stock price data were retrieved from Yahoo Finance for all firms in the sample [45]. Specifically, this study collected adjusted closing prices, which account for corporate actions such as stock splits, dividends, and capital gain distributions. These adjusted values provide a more accurate basis for analyzing shareholder returns over time compared to unadjusted prices.

Stock prices were aggregated on a daily basis and used to compute monthly average returns and volatility measures. To align with the temporal granularity of the sentiment and ESG variables, both daily and monthly aggregates were calculated for each firm. For firms headquartered outside the United States, currency normalization was performed using contemporaneous exchange rates. These rates—such as the EUR/USD exchange rate retrieved via the EUR=X ticker—were also obtained from *Yahoo Finance*. Foreign firms' stock prices were converted to USD equivalents to ensure comparability across firms and eliminate currency-related distortions in return calculations.

Additionally, ticker selection for multinational firms with dual listings was guided by data quality and liquidity. For instance, BNPQF was excluded in favor of BNPQY, due to the significantly higher trading volume of the latter. On December 30, 2020, BNPQY recorded a volume of 79,500 shares compared to just 1,100 shares for BNPQF, representing only 1.38% of BNPQY's liquidity. Using the more liquid ticker reduces noise and enhances the reliability of return-based analyses.

Among the various stock price transformations considered, this study selected the daily average of return-adjusted close, specifically the variable 'Adj Close Adjusted close price adjusted for splits and dividend and/or capital gain distributions._R_DAvg', as the primary dependent measure of stock performance. This variable captures the average day-over-day change in adjusted closing prices, smoothed over time to account for market closure days and daily volatility. Compared to alternative metrics such as raw adjusted prices (_DAvg) or unsmoothed returns (_R), 'Adj Close Adjusted close price adjusted for splits and dividend and/or capital gain distributions._R_DAvg' offers a more stable yet responsive indicator of firm-level performance. It reflects short-term investor reaction while maintaining robustness against outlier days or low-liquidity trading sessions.

**Table 3.5 Description of Stock Price Variables Derived from Adjusted Close Prices**

| Column Name | Description |
| --- | --- |
| Adj Close Adjusted close price adjusted for splits and dividend and/or capital gain distributions. | The closing price adjusted for stock splits, dividends, and/or capital gain distributions. |
| Adj Close Adjusted close price adjusted for splits and dividend and/or capital gain distributions.**_Daily_Avg** | The increase of the closing price adjusted for stock splits, dividends, and/or capital gain distributions per day (considering that market does not open everyday) |
| Adj Close Adjusted close price adjusted for splits and dividend and/or capital gain distributions.**_Daily_Avg_DAvg** | Daily average of the increase of the closing price adjusted for stock splits, dividends, and/or capital gain distributions per day (considering that market does not open everyday) |
| Adj Close Adjusted close price adjusted for splits and dividend and/or capital gain distributions.**_DAvg** | Daily average of the closing price adjusted for stock splits, dividends, and/or capital gain distributions. |
| Adj Close Adjusted close price adjusted for splits and dividend and/or capital gain distributions.**_R** | The increase of the closing price adjusted for stock splits, dividends, and/or capital gain distributions compared to previous market open day |
| Adj Close Adjusted close price adjusted for splits and dividend and/or capital gain distributions.**_R_DAvg** | Daily average of the increase of the closing price adjusted for stock splits, dividends, and/or capital gain distributions compared to previous market open day |

This choice was particularly important in the context of the ESG-sentiment interaction analysis, where alignment in temporal frequency and sensitivity to market movement is crucial. By using a normalized, daily return-based measure, this study ensured consistency in financial interpretation and enhanced comparability across firms and time horizons.

### 3.6 Feature Normalization

To ensure comparability across variables originating from heterogeneous sources—including ESG scores, retrofitted emotion similarity metrics, and financial return measures—all numerical features were normalized using min-max scaling. This method rescaled each variable to a fixed range of [0,1] using the transformation:

$$x' = \frac{x - x_{\min}}{x_{\max} - x_{\min}}$$

This choice was motivated by the bounded and interpretable nature of key features in the dataset, particularly cosine similarity scores and sentiment indicators derived from retrofitted word embeddings. Preserving their original distribution and scale was essential for ensuring a meaningful interpretation of regression coefficients, particularly when modeling interaction terms between sentiment, ESG ratings, and stock price variables. In this context, maintaining a uniform and bounded scale across all inputs prevents scale-driven biases and allows each feature to contribute comparably in statistical models.

In contrast, z-score normalization transforms variables to have a mean of zero and a standard deviation of one. While widely used in ESG rating construction to enable cross-firm benchmarking, z-score normalization introduces negative and unbounded values that may distort the interpretability of inherently bounded or skewed variables. Z-score normalization is also more appropriate when underlying distributions approximate normality or when preparing data for dimensionality reduction techniques such as Principal Component Analysis (PCA).

Several recent studies support the use of min-max normalization in ESG-integrated financial research. Deng et al. (2025) [46] found that ESG rating divergence can increase manufacturing firms' carbon emission intensity by inhibiting incentives for green innovation. Their analysis used min-max normalization to ensure interpretability and comparability across firms. Similarly, Benuzzi et al. (2025) [47] criticize percentile-based ESG normalization (e.g., in Refinitiv) for introducing distortions and recommend a simpler min-max approach to reflect true sustainability performance better. This aligns with findings from Kvam et al. (2024) [32], who also apply min-max normalization when analyzing the relationship between ESG sentiment and stock price reactions, ensuring consistent scaling for interpretability in regression-based models.

These findings collectively reinforce that min-max normalization is a theoretically sound and practically effective preprocessing strategy for regression-based ESG studies involving mixed data types. It enables interpretable interaction effects, accommodates bounded variables such as cosine similarities, and aligns well with the methodological needs of ESG–sentiment–finance integration.

**Table 3.6. Comparison of Min-Max and Z-Score Normalization**

| Criterion | Min-Max Normalization | Z-Score Normalization |
|---|---|---|
| **Definition** | Scales values to the range [0,1] | Standardizes values to mean = 0 and standard deviation = 1 |
| **Formula** | $x' = \dfrac{x - x_{\min}}{x_{\max} - x_{\min}}$ | $x' = \dfrac{x - \mu}{\sigma}$ |
| **Preserves Original Distribution** | Yes | No (alters distribution shape) |
| **Output Range** | Bounded between 0 and 1 | Unbounded |
| **Interpretability** | High (proportional and intuitive) | Moderate (relative to population mean and variance) |
| **Outlier Sensitivity** | High | Lower |
| **Use Case Suitability** | Ideal for bounded variables, interaction terms, and similarity metrics | Ideal for normally distributed features or PCA (Principal Component Analysis) preprocessing |
| **Compatibility with Cosine Similarity** | Excellent (retains original magnitude and range) | Risk of distortion in similarity-based modeling |

## 3.7 Triple Significance Criteria

$$\text{ESASR}_{i,t} = \alpha + \beta_1 \cdot \text{ESG}_{i,t} + \beta_2 \cdot \text{Sentiment}_{i,t} + \beta_3 \cdot \left( \text{ESG}_{i,t} \times \text{Sentiment}_{i,t} \right) + \varepsilon_{i,t}$$

Triple Significance Criteria To improve the rigor of the empirical analysis, this study introduced a "triple significance" filter. This criterion required that the ESG score, sentiment variable, and their interaction term all be statistically significant at the 10% level (p < 0.1) within the same regression model. The rationale behind this approach was to isolate cases where sentiment not only independently correlates with stock returns but also meaningfully moderates the effect of ESG performance. This filtering step ensures that only robust and jointly influential relationships are emphasized in the interpretation of results.

## 3.8 Comparative Evaluation of Sentiment Modeling Techniques: Retrofitted Word Embeddings vs. NRC Emotion Lexicon

**Table 3.8.a. Overview of Sentiment Variable Substitutions for Comparative Analysis**

| Column Name | Description | Replaced Column |
|---|---|---|
| InpageTitle_NRC_joy | Count of joy-related words in the yearly collective title via NRC Emotion Lexicon | InpageTitle_Retro_happy_similarity |
| InpageTitle_NRC_sadness | Count of sadness-related words in the yearly collective title via NRC Emotion Lexicon | InpageTitle_Retro_sad_similarity |
| InpageTitle_NRC_anger | Count of anger-related words in the yearly collective title via NRC Emotion Lexicon | InpageTitle_Retro_angry_similarity |
| InpageTitle_NRC_fear | Count of fear-related words in the yearly collective title via NRC Emotion Lexicon | InpageTitle_Retro_fear_similarity |
| InpageTitle_NRC_surprise | Count of surprise-related words in the yearly collective title via NRC Emotion Lexicon | InpageTitle_Retro_surprise_similarity |
| InpageTitle_NRC_trust | Count of trust-related words in the yearly collective title via NRC Emotion Lexicon | InpageTitle_Retro_trust_similarity |
| InpageTitle_NRC_disgust | Count of disgust-related words in the yearly collective title via NRC Emotion Lexicon | InpageTitle_Retro_disgust_similarity |
| InpageTitle_NRC_anticipation | Count of anticipation-related words in the yearly collective title via NRC Emotion Lexicon | InpageTitle_Retro_anticipation_similarity |
| InpageTitle_NRC_joy_DAvg | Daily average of count of joy-related words in the title via NRC Emotion Lexicon | InpageTitle_Retro_happy_similarity_DAvg |
| InpageTitle_NRC_sadness_DAvg | Daily average of count of sadness-related words in the title via NRC Emotion Lexicon | InpageTitle_Retro_sad_similarity_DAvg |
| InpageTitle_NRC_anger_DAvg | Daily average of count of anger-related words in the title via NRC Emotion Lexicon | InpageTitle_Retro_angry_similarity_DAvg |
| InpageTitle_NRC_fear_DAvg | Daily average of count of fear-related words in the title via NRC Emotion Lexicon | InpageTitle_Retro_fear_similarity_DAvg |
| InpageTitle_NRC_surprise_DAvg | Daily average of count of surprise-related words in the title via NRC Emotion Lexicon | InpageTitle_Retro_surprise_similarity_DAvg |
| InpageTitle_NRC_trust_DAvg | Daily average of count of trust-related words in the title via NRC Emotion Lexicon | InpageTitle_Retro_trust_similarity_DAvg |
| InpageTitle_NRC_disgust_DAvg | Daily average of count of disgust-related words in the title via NRC Emotion Lexicon | InpageTitle_Retro_disgust_similarity_DAvg |
| InpageTitle_NRC_anticipation_DAvg | Daily average of count of anticipation-related words in the title via NRC Emotion Lexicon | InpageTitle_Retro_anticipation_similarity_DAvg |

Table 3.8.a presents the specific sentiment-related variables that were substituted to facilitate a comparative evaluation of two sentiment modeling techniques: Retrofitted Word Embeddings and the NRC Emotion Lexicon.

In this comparative framework, the original variables—quantifying semantic similarity to emotion-laden words via retrofitted embeddings—were replaced with their counterparts derived from the NRC Emotion Lexicon, which rely on word counts associated with predefined emotional categories. The analysis was conducted using an identical empirical design and model specifications across both approaches. The only difference was the set of sentiment variables listed in Table 3.8.a. Both aggregate yearly and daily average sentiment measures were examined across a consistent textual corpus (yearly in-page titles), enabling a controlled assessment of the relative efficacy of the two sentiment quantification methodologies within the context of ESG-related textual analysis.

# 4. Results

## 4.1 Descriptive Statistics Table

**Table 4.1a.. Descriptive Statistics (Not Normalized, N = 209)**

|  | mean | std | min | max |
|---|---|---|---|---|
| overall_score | 0.7613 | 0.2512 | 0.0323 | 0.9968 |
| econ_score | 0.6270 | 0.3025 | -0.0788 | 1.1723 |
| envrn_score | 0.8066 | 0.2591 | -0.1501 | 1.0934 |
| corpgov_score | 0.7176 | 0.2495 | 0.0556 | 0.9962 |
| social_score | 0.7000 | 0.2641 | 0.0430 | 0.9720 |
| InpageTitle_Retro_happy_similarity | 0.1532 | 0.0539 | -0.0682 | 0.2508 |
| InpageTitle_Retro_sad_similarity | 0.1127 | 0.0247 | 0.0250 | 0.1598 |
| InpageTitle_Retro_angry_similarity | 0.1573 | 0.0388 | -0.0928 | 0.2238 |
| InpageTitle_Retro_fear_similarity | 0.1600 | 0.0470 | -0.1184 | 0.2696 |
| InpageTitle_Retro_surprise_similarity | 0.0721 | 0.0288 | -0.0410 | 0.1554 |
| InpageTitle_Retro_trust_similarity | 0.3287 | 0.0437 | 0.0604 | 0.3942 |
| InpageTitle_Retro_disgust_similarity | -0.1835 | 0.0261 | -0.2208 | -0.0425 |
| InpageTitle_Retro_anticipation_similarity | 0.4649 | 0.0564 | 0.1051 | 0.6039 |
| InpageTitle_Retro_happy_similarity_DAvg | 0.1213 | 0.0393 | 0.0073 | 0.2229 |
| InpageTitle_Retro_sad_similarity_DAvg | 0.0889 | 0.0225 | 0.0348 | 0.1462 |
| InpageTitle_Retro_angry_similarity_DAvg | 0.1237 | 0.0336 | 0.0129 | 0.1955 |
| InpageTitle_Retro_fear_similarity_DAvg | 0.1261 | 0.0393 | 0.0093 | 0.2724 |
| InpageTitle_Retro_surprise_similarity_DAvg | 0.0558 | 0.0237 | -0.0016 | 0.1327 |
| InpageTitle_Retro_trust_similarity_DAvg | 0.2610 | 0.0456 | 0.0602 | 0.3446 |
| InpageTitle_Retro_disgust_similarity_DAvg | -0.1476 | 0.0253 | -0.1881 | -0.0056 |
| InpageTitle_Retro_anticipation_similarity_DAvg | 0.3702 | 0.0501 | 0.1642 | 0.4710 |
| Adj Close Adjusted close price adjusted for splits and dividend and/or capital gain distributions._R_DAvg | 1.0053 | 0.0436 | 0.8526 | 1.5744 |

**Table 4.1.b. Descriptive Statistics (Normalized, N = 209)**

|  | mean | std | min | max |
|---|---|---|---|---|
| overall_score | 0.5918 | 0.3324 | 0.0 | 1.0 |
| econ_score | 0.6036 | 0.3458 | 0.0 | 1.0 |
| envrn_score | 0.5303 | 0.3127 | 0.0 | 1.0 |
| corpgov_score | 0.5354 | 0.3367 | 0.0 | 1.0 |
| social_score | 0.5371 | 0.3275 | 0.0 | 1.0 |
| InpageTitle_Retro_happy_similarity | 0.5057 | 0.3224 | 0.0 | 1.0 |
| InpageTitle_Retro_sad_similarity | 0.4904 | 0.3222 | 0.0 | 1.0 |
| InpageTitle_Retro_angry_similarity | 0.5456 | 0.3161 | 0.0 | 1.0 |
| InpageTitle_Retro_fear_similarity | 0.5170 | 0.3108 | 0.0 | 1.0 |
| InpageTitle_Retro_surprise_similarity | 0.4749 | 0.3263 | 0.0 | 1.0 |
| InpageTitle_Retro_trust_similarity | 0.5552 | 0.3175 | 0.0 | 1.0 |
| InpageTitle_Retro_disgust_similarity | 0.4190 | 0.3139 | 0.0 | 1.0 |
| InpageTitle_Retro_anticipation_similarity | 0.5492 | 0.3167 | 0.0 | 1.0 |
| InpageTitle_Retro_happy_similarity_DAvg | 0.4396 | 0.3141 | 0.0 | 1.0 |
| InpageTitle_Retro_sad_similarity_DAvg | 0.4822 | 0.3110 | 0.0 | 1.0 |
| InpageTitle_Retro_angry_similarity_DAvg | 0.5233 | 0.3163 | 0.0 | 1.0 |
| InpageTitle_Retro_fear_similarity_DAvg | 0.4602 | 0.3009 | 0.0 | 1.0 |
| InpageTitle_Retro_surprise_similarity_DAvg | 0.4487 | 0.3272 | 0.0 | 1.0 |
| InpageTitle_Retro_trust_similarity_DAvg | 0.5534 | 0.3149 | 0.0 | 1.0 |
| InpageTitle_Retro_disgust_similarity_DAvg | 0.4267 | 0.3157 | 0.0 | 1.0 |
| InpageTitle_Retro_anticipation_similarity_DAvg | 0.5402 | 0.3157 | 0.0 | 1.0 |
| Adj Close Adjusted close price adjusted for splits and dividend and/or capital gain distributions._R_DAvg | 0.4483 | 0.3005 | 0.0 | 1.0 |

Table 4.1.a and Table 4.1.b present summary statistics for all variables used in the study, including ESG scores, emotion-specific sentiment similarities derived from retrofitted word embeddings, and adjusted stock return data. Both the original (non-normalized) and min-max normalized values are reported to illustrate scale differences and the need for transformation.

To ensure comparability across features with different scales and distributions, all variables were normalized to a [0,1] range using min-max scaling. This method preserves original distributions and is particularly suitable for bounded metrics like cosine similarities. It also facilitates the interpretation of interaction terms in regression models, avoiding the distortion risks associated with z-score standardization. Post-normalization, variable rankings and spread remain consistent, supporting reliable analysis of ESG–sentiment–return dynamics.

## 4.2 Correlation Table

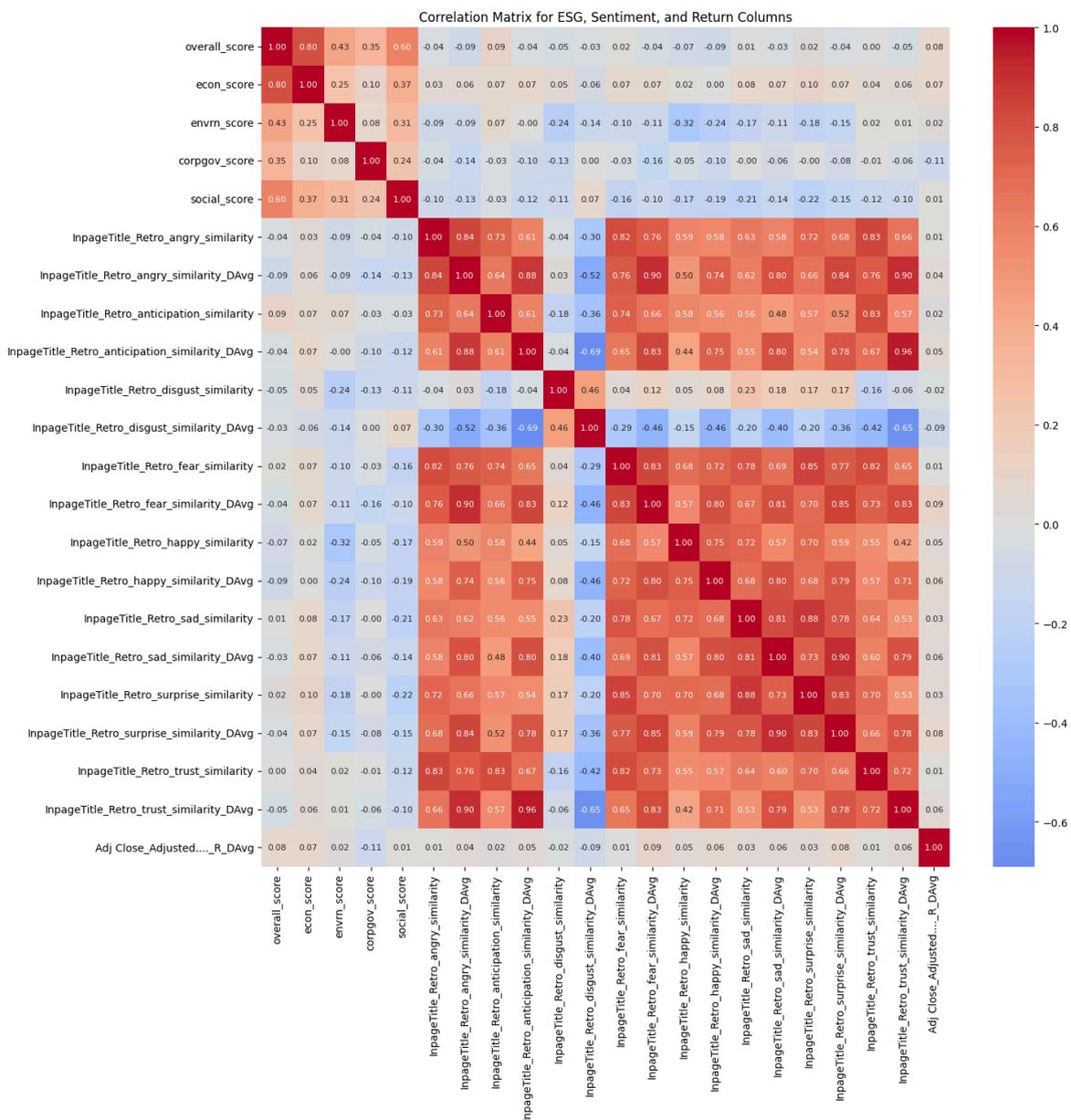

**Figure 4.2 Correlation Matrix Between ESG Scores and Sentiment Dimensions**

Figure 4.2 displays a heatmap illustrating the pairwise correlations among ESG sub-scores (economic, environmental, social, and governance) and sentiment similarity scores derived from retro-themed content. A clear pattern of positive correlations emerges both within the ESG categories and within the sentiment dimensions, indicating internal coherence and thematic consistency in each group. Notably, the environmental score exhibits the strongest and most consistent positive associations with various sentiments, underscoring its central role in the sentiment-ESG relationship.

While most sentiment dimensions—such as trust, anticipation, and surprise—are positively correlated with ESG scores, 'disgust' stands out as an exception, showing a negative correlation across several ESG components. This divergence suggests that 'disgust' may reflect market or perceptual disapproval, in contrast to the more affirmative tone conveyed by other sentiments. The environmental dimension's prominence in its significant and broad correlations with sentiments further emphasizes its potential as a mediating variable in sentiment-driven ESG evaluations.

## 4.3 Result of Comparative Evaluation of Sentiment Modeling Techniques: Retrofitted Word Embeddings vs. NRC Emotion Lexicon

**Table 4.3.1. Comparison of Retrofitted vs. NRC Sentiment Models on Triple-Significant ESG × Sentiment Interactions**

| Metric | Retrofitted Word Embeddings | NRC Emotion Lexicon |
|---|---|---|
| Average $R^2$ (Triple-Significant Only) | 0.619 | 0.572 |
| Average $R^2$ (All Interactions) | 0.301 | 0.273 |
| Total Triple-Significant Interactions (n) | 85 | 90 |

The comparative analysis between Retrofitted Word Embeddings and NRC Emotion Lexicon sentiment methods reveals notable differences in their predictive effectiveness. Specifically, Retrofitted Word Embeddings sentiment interactions that were statistically significant across all three dimensions (ESG × Sentiment × Return) exhibited a higher average $R^2$ of 0.619 compared to NRC Emotion Lexicon's 0.572. When considering all interactions, Retrofitted Word Embeddings still maintained a higher average explanatory power ($R^2 = 0.301$) relative to NRC Emotion Lexicon ($R^2 = 0.273$). Although the total number of triple-significant interactions was slightly lower for Retrofitted Word Embeddings (85) than NRC Emotion Lexicon (90), the consistently higher $R^2$ values support Retrofitted Word Embeddings' greater explanatory power in the ESG–sentiment interaction context. Therefore, this analysis justifies the use of Retrofitted Word Embeddings sentiment as a more robust and insightful tool for capturing sentiment-driven ESG effects on stock returns.

## 4.4 Reasoning of Dual Sentiment Aggregation Strategies: _Davg Sentiment and Non-_Davg Sentiment

To robustly capture the emotional tone in financial narratives, this study employs two complementary sentiment aggregation strategies: (1) non-_Davg sentiment variables, calculated by pooling all headlines across the year, and (2) _Davg sentiment variables, constructed through a two-tier averaging process—first at the daily level and then across the year. This dual approach addresses a key methodological concern in sentiment analysis: balancing data completeness with temporal normalization.

Table 3.3.b summarizes the core differences between these approaches. Non-_Davg sentiment variables are sensitive to article volume and can overrepresent periods of high news intensity. While this makes them useful for capturing overall emotional exposure, they may also introduce bias if media coverage is unevenly distributed throughout the year. In contrast, _Davg sentiment variables mitigate such bias by ensuring each day contributes equally to the final annual score. This makes them better suited for isolating the consistent, time-distributed emotional climate associated with each firm.

Empirical performance further supports the value of using both approaches. Among regression models meeting triple significance, sentiment variables based on _Davg sentiment appeared more frequently—accounting for 49 models—compared to 36 models using non-_Davg sentiments. On the other hand, the models using non-_Davg sentiments showed slightly higher average explanatory power (mean R² = 0.629) than their _Davg sentiment(mean R² = 0.613).

When considering the full model set without filtering for significance (680 models per group), _Davg sentiments again demonstrated a slight advantage, achieving a higher average R² (0.287 vs. 0.281). These results underscore the strengths and trade-offs of each method. The _Davg sentiments yield more consistent signals by normalizing temporal variation, while the non-_Davg sentiments may better capture effects related to information intensity or news clustering. Employing both allows for a richer analysis of how sentiment operates within ESG–return dynamics, from both a stable daily-tone perspective and a cumulative exposure viewpoint.

## 4.5 Overview of Model Significance

This study estimates firm-level panel regressions to evaluate how ESG ratings, headline-level emotion-specific sentiment, and their interaction influence short-term stock returns. Significance was assessed at the 10% level across models, and adjusted R² values were considered to evaluate explanatory power. Models were estimated separately for each firm using normalized variables and included interaction terms to test moderation effects as posited in Hypotheses 1–3.

## 4.6 H1: Emotion-Specific Sentiment Triple Significance and Stock Returns

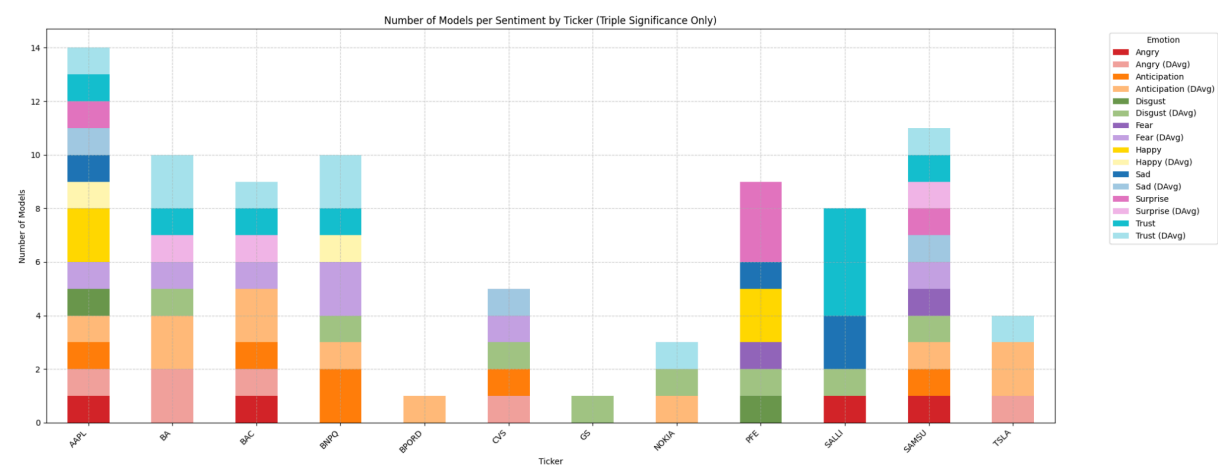

Figure 4.6.a. Distribution of Emotion-Specific Triple Significance Models Across Companies (Triple-Significance Models Only)

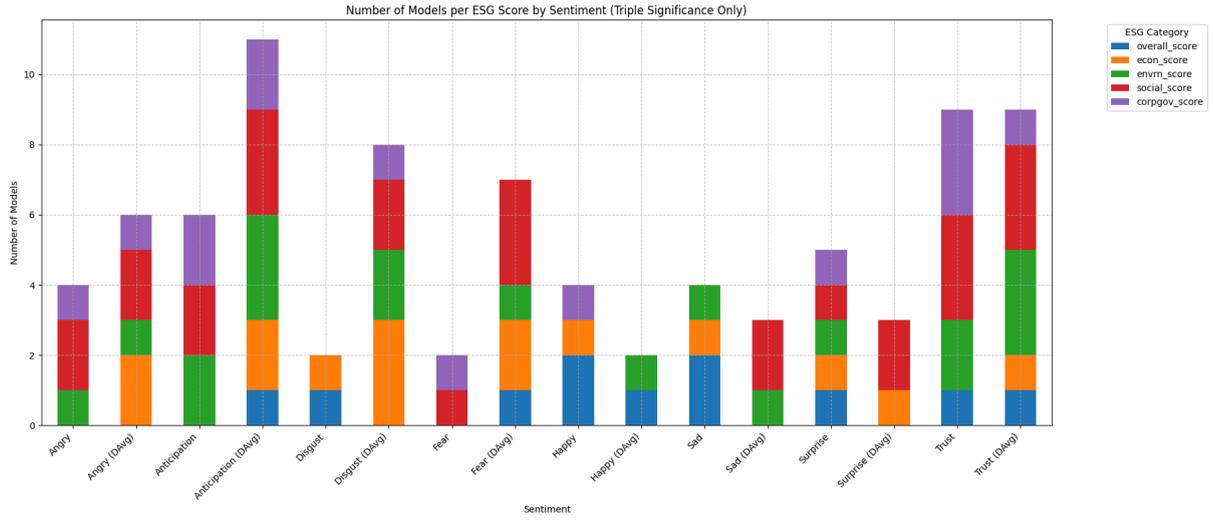

**Figure 4.6.b. Distribution of Triple Significance Models by Emotion Type and ESG Dimension (Triple-Significance Models Only)**

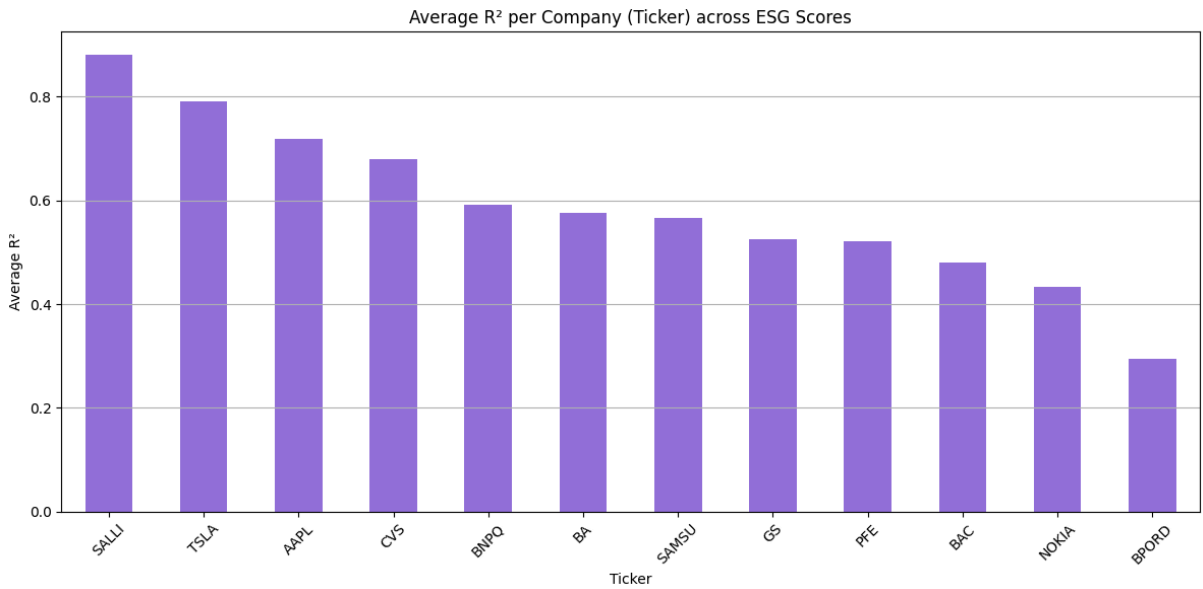

**Figure 4.6.c. Average R² per Company from Sentiment × ESG Interaction Models (Triple-Significance Models Only)**

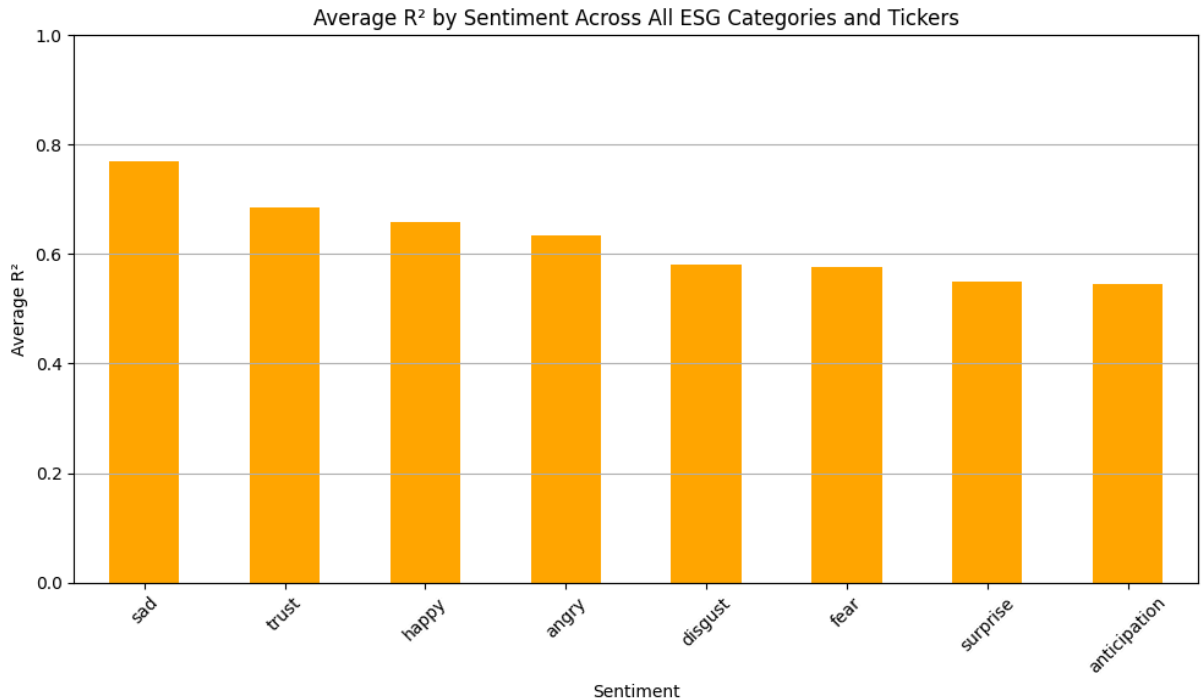

**Figure 4.6.d. Average R² by Sentiment Across All ESG Categories and Tickers (Triple-Significance Models Only)**

To evaluate Hypothesis 1, this study examined whether emotion-specific sentiment signals embedded in financial news headlines significantly influence stock returns when contextualized by firm-level ESG performance. Regression framework incorporated individual emotion similarity metrics, ESG scores, and their interaction term (ESG_x_Sentiment). A model was deemed supportive of H1 only when all three components were statistically significant ($p < 0.1$)—a condition referred to as triple significance.

Out of 16 companies, 12 exhibited at least one model meeting this criterion, producing a total of 98 triple significance observations. This provides compelling support for H1, indicating that sentiment—when aligned with ESG context—exerts a meaningful effect on financial market outcomes.

Emotion diversity was also pronounced. For example, AAPL exhibited the highest number of triple significance models (14), spanning all eight core emotions and their daily-averaged (_DAvg) counterparts. Other tickers such as BA, BNPQ, and SAMSU also demonstrated a rich emotional spread, each with models linked to at least seven different sentiment types. Notably, PFE and SALLI showed concentrated emotion diversity, with strong representation from sentiments like *Surprise* and *Trust*. This distribution of emotion-specific triple significance models is visualized in Figure 4.6.a.

Among all metrics, *Anticipation_Davg* emerged most frequently (11 times), followed by *Trust* and *Trust_Davg* (9 each), and *Disgust_Davg* (8 times), underscoring the nuanced influence of discrete emotions in ESG-aware financial modeling as shown in Figure 4.6.b

The robustness of these models is further confirmed by their explanatory power. As shown in Figure 4.6.c, several firms, including SALLI, TSLA, and AAPL, achieved average R² values exceeding 0.7, with some individual models exceeding 0.9. These high values affirm that sentiment-enhanced ESG models capture substantial return variation, validating the core claim of H1.

In terms of sentiment types, Figure 4.6.d indicates that emotions like Sad, Trust, and Happy produce higher average R² values, often above 0.6, highlighting their particular relevance in financial decision contexts. These findings confirm that emotion-specific sentiment signals significantly enhance return prediction when integrated with ESG performance.

## 4.7 H2: ESG Dimension-Specific Moderation Effects

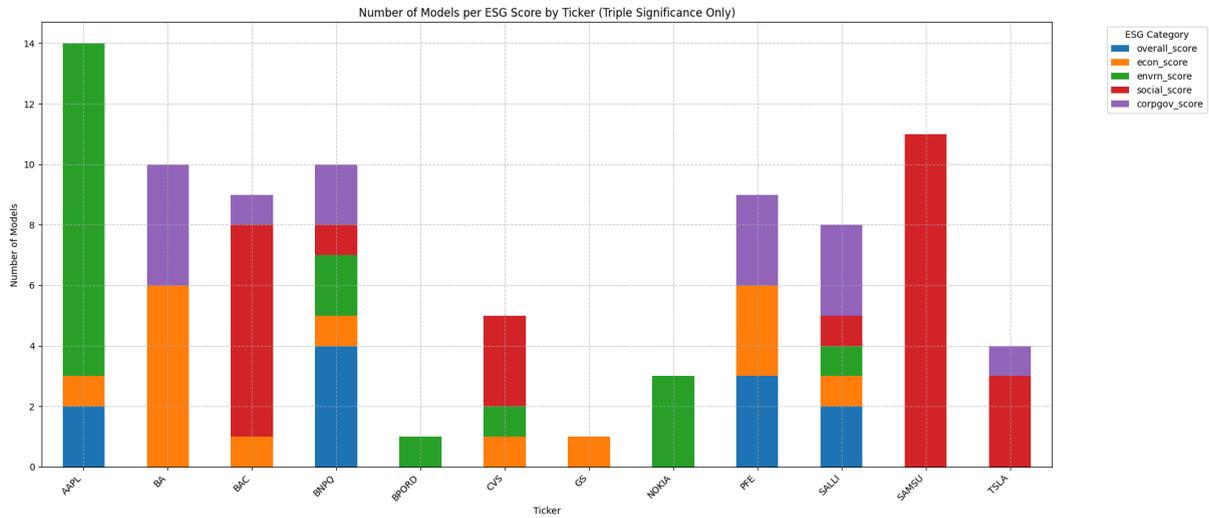

**Figure 4.7.a. Distribution of Triple Significance Models by ESG Dimension Across Companies (Triple-Significance Models Only)**

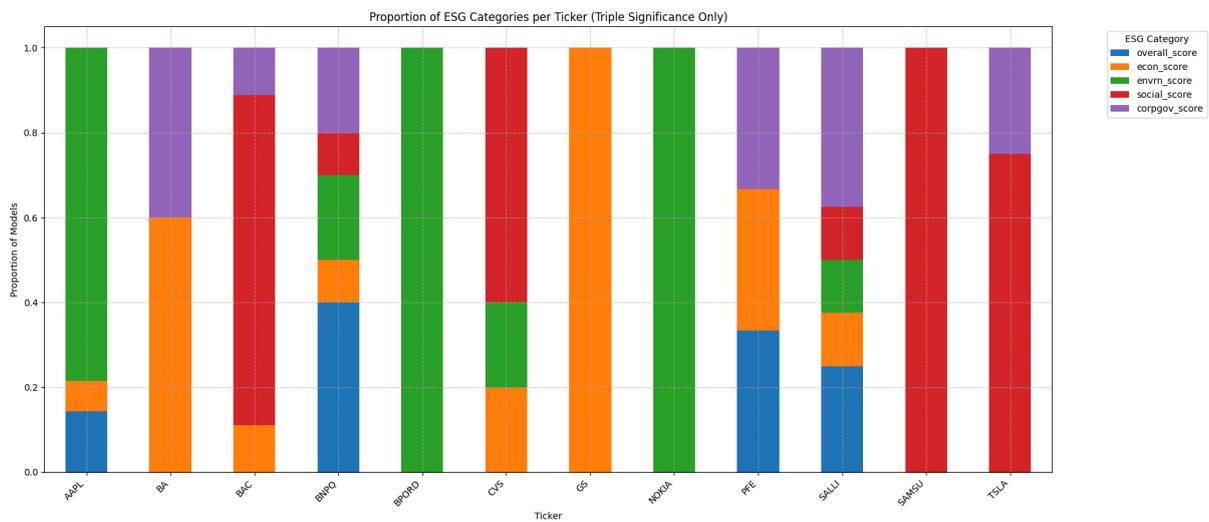

**Figure 4.7.b. Proportional Distribution of ESG Category Associations per Ticker (Triple-Significance Models Only)**

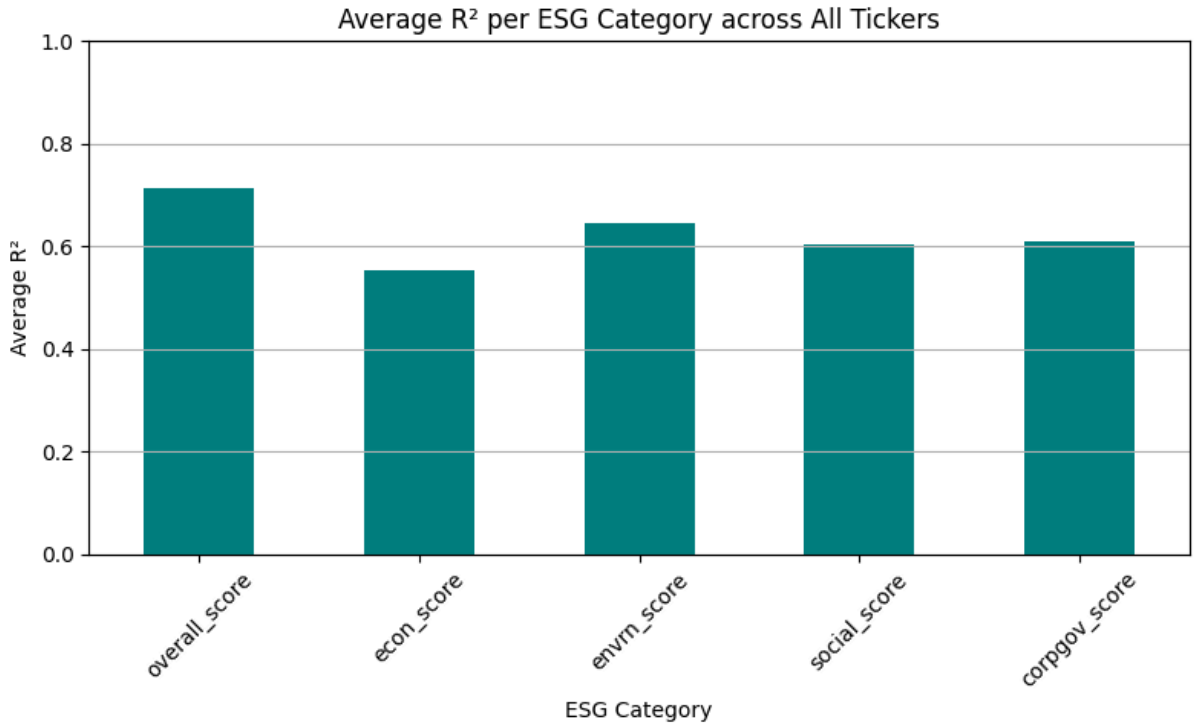

Figure 4.7.c. Average R² by ESG Category Across All Tickers (Triple-Significance Models Only)

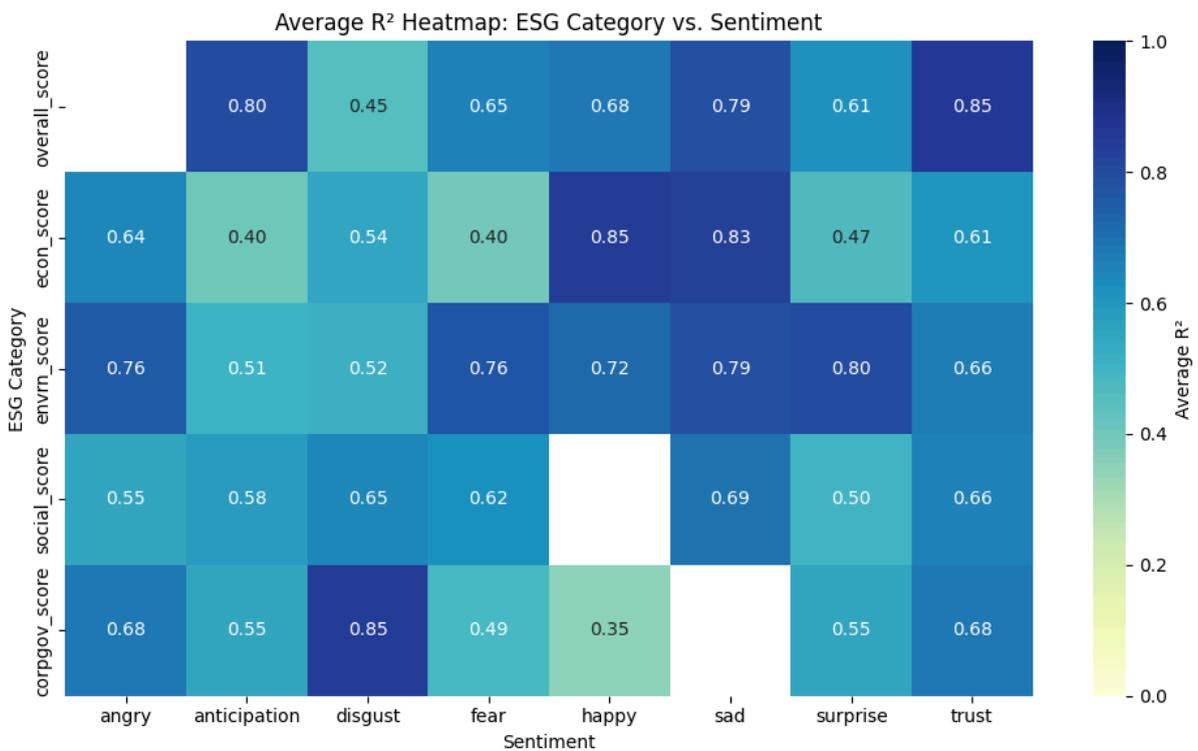

Figure 4.7.d. Average R² Heatmap: ESG Category vs. Sentiment (Triple-Significance Models Only)

To evaluate Hypothesis 2, this study examined whether the moderating effect of emotion-specific sentiment on ESG-related stock returns varies across ESG subdimensions: Environmental (E), Social (S), and Governance (G). The results provide strong evidence in support of this hypothesis, revealing that both the presence and strength of sentiment effects differ significantly by ESG category.

First, Figure 4.7.a and Figure 4.7.b show that companies tend to cluster around specific ESG dimensions in their triple significance models. For example, SAMSU, TSLA, BAC, and CVS exhibited significance almost exclusively within the Social dimension, whereas AAPL, NOKIA, and BPORD were more prominent in Environmental interactions. This points to firm-level ESG salience, where investors likely prioritize different ESG pillars depending on industry or disclosure norms.

Second, Figure 4.6.b visualizes how emotion types—*Anticipation, Trust,* and *Disgust* in particular—distribute across ESG dimensions, further emphasizing that the sentiment-ESG interaction is dimension-specific. These findings collectively confirm H2, illustrating that emotion-driven market responses are shaped not only by sentiment type but also by the ESG content being evaluated.

Third, Figure 4.7.c summarizes average $R^2$ values across ESG categories and indicates that models involving overall ESG performance (overall_score) and environmental factors (envrn_score) exhibit the highest explanatory power, with mean $R^2$ values exceeding 0.6. Conversely, economic (econ_score), social (social_score), and corporate governance (corpgov_score) categories show comparatively lower explanatory power, suggesting these domains may be less directly reflected in sentiment-driven pricing behaviors within this sample. These findings reinforce the notion that sentiment impacts ESG dimensions differently, likely depending on how tangible or measurable each category's outcomes appear to investors.

Lastly, further nuance is captured in Figure 4.7.d, which maps $R^2$ values by sentiment type and ESG dimension. The highest explanatory power is observed in the combinations of corpgov_score × disgust ($R^2 = 0.85$), overall_score × trust ($R^2 = 0.85$), econ_score × happy ($R^2 = 0.85$), and econ_score × sad ($R^2 = 0.83$). This heatmap highlights not only the ESG categories that are more sensitive to sentiment modulation but also specifies which emotional tones (e.g., disgust, trust, happy, sad) yield the greatest influence within each ESG domain.

Together, these findings confirm H2: the effect of sentiment on ESG-related stock return prediction is not uniform across ESG dimensions. The emotional salience of specific ESG categories varies by firm and investor perception, with some dimensions (like Environmental and Economic) being more immediately translated into market behavior when sentiment is factored in.

## 4.8 H3: Sentiment as a Moderator of ESG–Return Relationships: Empirical Results

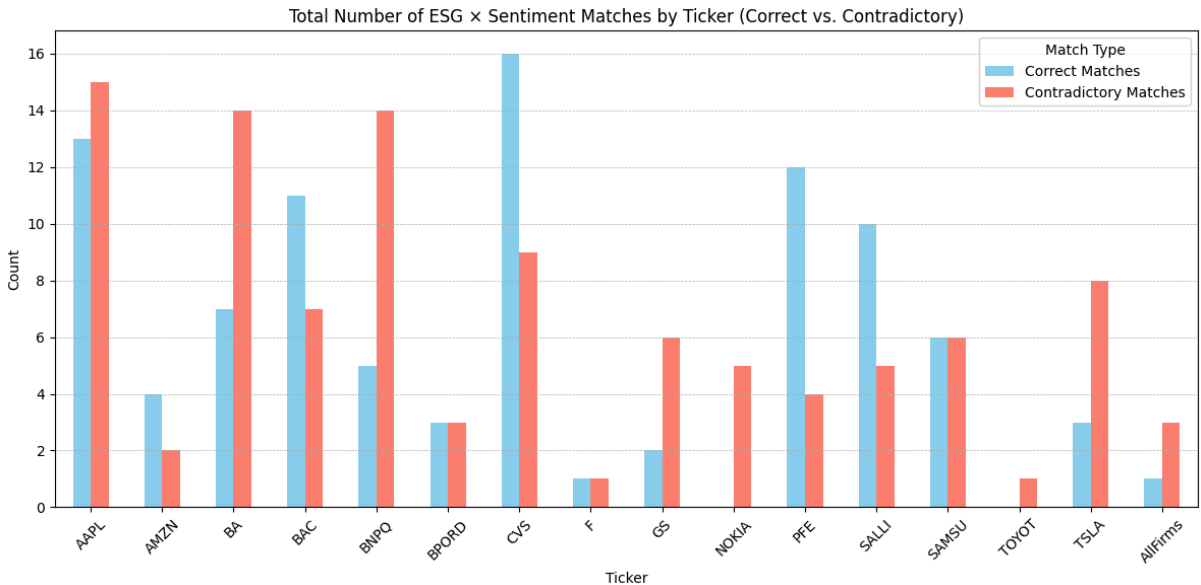

**Figure 4.8.1. ESG × Sentiment Interaction Counts (Correct vs. Contradictory) Across Firms**

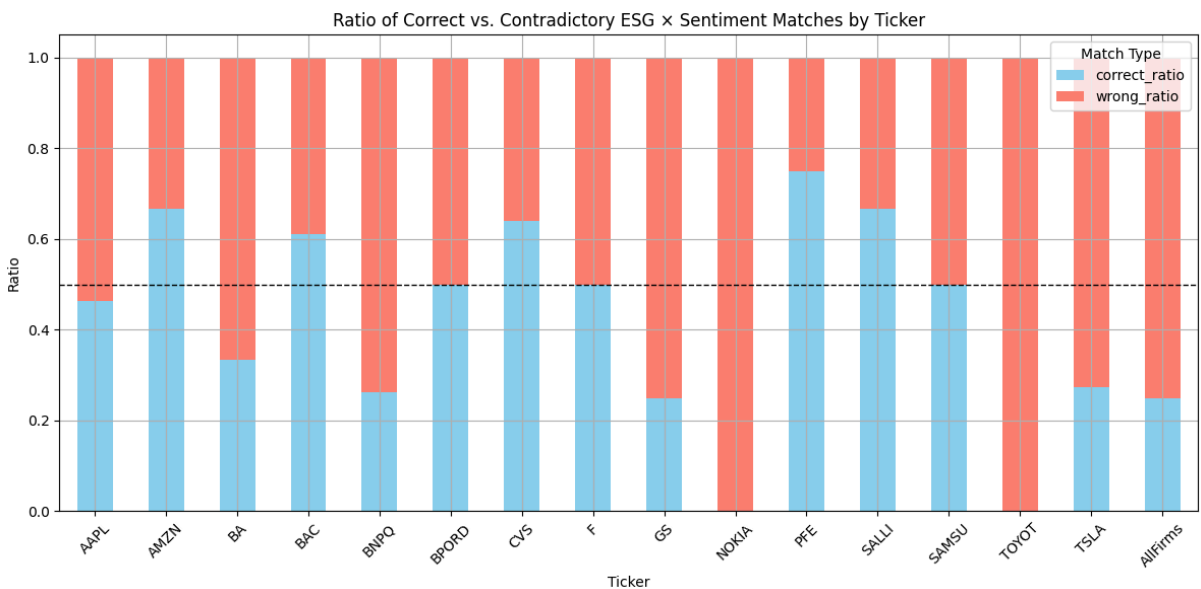

**Figure 4.8.2. Correct and Contradictory ESG × Sentiment Interaction Ratios Across Firms**

Hypothesis 3 posits that the relationship between ESG performance and stock returns is moderated by sentiment, specifically, that positive sentiment amplifies and negative sentiment dampens the financial impact of ESG scores. To evaluate this, this study analyzed 220 statistically significant ESG × Sentiment interaction terms ($p < 0.1$) across multiple firms.

The analysis categorized sentiment into three distinct groups: positive (trust, anticipation, happy), neutral (surprise), and negative (disgust, angry, fear, sad). Each sentiment type was evaluated in terms of its interaction with ESG performance across multiple firms. The findings reveal that positive sentiments generally amplify the relationship between ESG scores and stock returns, whereas negative sentiments tend to diminish this relationship.

Since neutral sentiment does not directly relate to the hypothesis, 27 significant interaction terms involving "surprise" were excluded from further analysis.

Out of the remaining 197 statistically significant ESG × Sentiment interactions, 94 interactions (47.7%) supported Hypothesis 3, meaning the direction of the sentiment (positive or negative) matched the sign of the coefficient. Conversely, 103 interactions (52.3%) contradicted H3.

Figure 4.8.1 displays the distribution of correct and contradictory matches across firms, while Figure 4.8.2 illustrates their proportions. Companies like AMZN, BAC, CVS, PFE, SALLI exhibited strong alignment with H3, while firms such as NOKIA, TOYOT, GS, TSLA, BNPQ, BA had a higher share of contradictory interactions. Notably, the statistically significant ESG × Sentiment interactions in the 'AllFirms' dataset demonstrated a pronounced tendency toward contradictory interactions, indicating an overall weaker support for Hypothesis 3 across the aggregated market.

$$\text{Weighted Avg} = \frac{\sum \left( \frac{1}{SE_i^2} \cdot \beta_i \right)}{\sum \left( \frac{1}{SE_i^2} \right)}$$

$\beta_i$: denotes the estimated coefficient

$SE_i$: the standard error.

To assess the net effect size of sentiment moderation, this study computed the weighted average of all 197 ESG × Sentiment coefficients, using inverse standard errors as weights. The resulting weighted average across all significant interactions was 0.179 (p-value < 0.0001), indicating that, on average, sentiment modestly enhances the financial impact of ESG performance.

However, when isolating only the subset of 94 interactions consistent with Hypothesis 3, the weighted average coefficient was negative (-0.215; p-value = 0.0005). This implies that even interactions aligning with the expected direction (e.g., positive sentiment amplifying positive ESG impacts) yield an overall suppressive effect.

Interestingly, for the 103 interactions contradicting Hypothesis 3, the weighted average coefficient was significantly positive at 0.541 (p-value < 0.0001). Thus, contradictory interactions appear to exhibit a stronger enhancing effect, suggesting complex dynamics between ESG performance, sentiment, and stock returns that differ from initial theoretical expectations.

Table 4.8.1. Weighted Average ESG × Sentiment Interaction Effects by H3 Alignment

| Interaction Subset | No. of Interactions | Weighted Avg. Coefficient | p-value | Interpretation |
|---|---|---|---|---|
| All Significant Interactions | 197 | 0.179 | < 0.0001 | Sentiment modestly enhances ESG impact |
| H3-Consistent Interactions | 94 | -0.215 | 0.0005 | Net suppressive effect despite expected direction |
| H3-Contradictory Interactions | 103 | 0.541 | < 0.0001 | Strong enhancing effect, contrary to H3 |

These findings present a nuanced picture of Hypothesis 3, offering mixed and somewhat contradictory support. Sentiment clearly moderates the ESG–return relationship; however, contrary to initial expectations, interactions consistent with the hypothesis exhibited an overall suppressive effect. Interestingly, contradictory interactions demonstrated a notably stronger enhancing effect. This pattern suggests that the market response to ESG-related sentiment is more complex than anticipated, with positive sentiment not necessarily amplifying, nor negative sentiment uniformly dampening, ESG impacts on stock returns.

## 5. Discussion

This study provides new insights into the interplay between ESG performance and financial news sentiment, demonstrating that emotion-specific sentiment significantly affects how markets interpret ESG signals. By leveraging retrofitted word embeddings to extract nuanced emotional cues from financial headlines, this study uncovers a layered behavioral dimension in ESG investing.

First, the evidence strongly supports Hypothesis 1: sentiment, when captured with emotion specificity and contextualized through ESG performance, significantly predicts short-term stock returns. The triple-significance filtering approach revealed that certain emotions—such as Trust, Anticipation, and Sad—exert consistently strong effects on return outcomes. This confirms prior behavioral finance literature suggesting that emotions in financial narratives can shape market movements and investor decision-making.

Second, the study validates Hypothesis 2 by demonstrating that sentiment moderation effects vary across ESG subdimensions. Environmental and Economic scores generally displayed higher explanatory power, while Social and Governance components were less consistently aligned with sentiment-modulated return predictions. This finding suggests that investors respond differently to ESG information depending on the perceived tangibility or materiality of that ESG category, a behavior consistent with industry- and sector-specific risk salience.

However, the results for Hypothesis 3 reveal a more complex picture. While 47.7% of ESG × Sentiment interactions followed the expected direction—positive sentiment enhancing and negative sentiment dampening ESG effects—the weighted average coefficient among these H3-consistent interactions was negative. In contrast, contradictory interactions yielded a significantly positive average effect. These results challenge the symmetry implied in H3 and suggest that negative sentiment may possess stronger market influence than positive cues, or that markets respond differently to sentiment when expectations are not aligned. This may reflect investor bias, media negativity effects, or asymmetric reactions to risk signals.

Furthermore, the comparison between sentiment modeling methods supports the decision to use retrofitted word embeddings. Retrofitted sentiment consistently outperformed the NRC Emotion Lexicon across both overall and triple-significant interactions, justifying its application for fine-grained sentiment analysis in ESG contexts.

Taken together, these findings emphasize that sentiment—particularly when measured with emotion specificity—serves as a vital moderator in the ESG–return relationship. The complexity observed in Hypothesis 3 results suggests that future models should account for asymmetries in sentiment influence, perhaps incorporating non-linear terms or time-sensitive dynamics. These insights can inform both academic research and practical ESG investing strategies, urging a more behavioral and dynamic approach to ESG evaluation.

# 6. Implications and Contributions

This study makes several key contributions to the ESG and behavioral finance literature, while also offering practical implications for investors, analysts, and policymakers.

## 6.1 Theoretical Contributions

Behavioral Integration into ESG Models: By incorporating emotion-specific sentiment derived from retrofitted word embeddings, this research advances traditional ESG analysis through a behavioral lens. It moves beyond static ESG scores to explore how public perception, as captured in financial headlines, dynamically moderates ESG impacts on returns.

Disaggregation of Sentiment Effects: The findings highlight the importance of distinguishing among emotional tones (e.g., trust, fear, anticipation) rather than relying on broad positive/negative sentiment classifications [48]. This granularity enables more precise modeling of market reactions and adds new depth to behavioral asset pricing theories.

## 6.2 Empirical Contributions

Triple Significance Filtering: The introduction of the "triple significance" criterion ensures that the interaction effects reported are not incidental, but jointly meaningful. This methodological innovation enhances robustness and provides a replicable standard for future ESG-sentiment interaction studies.

Comparative Validation of Sentiment Models: The study empirically demonstrates that retrofitted sentiment embeddings provide greater explanatory power than lexicon-based approaches (e.g., NRC), reinforcing the value of emotion-enriched NLP techniques in financial analysis.

## 6.3 Practical Implications

Investor Strategy Enhancement: For asset managers and analysts, incorporating emotion-specific sentiment as a filtering or weighting mechanism can improve ESG-integrated investment decisions. The results point to emotion types like trust and anticipation as particularly valuable indicators of market behavior.

Policy and Disclosure Guidance: Regulators and sustainability disclosure bodies may benefit from understanding how media sentiment influences ESG interpretation. Crafting clearer narratives around ESG practices can help reduce market misinterpretations and align firm value with long-term sustainability goals.

By bridging ESG metrics, emotional sentiment, and financial returns, this study provides a comprehensive framework for capturing the psychological drivers of ESG investing, contributing meaningfully to both academic literature and real-world financial practice.

# 7. Limitations and Future Research Directions

## 7.1 Time-focused Limitations and Directions

A primary limitation of this study lies in the sample size. Only 13 observations were available for the final model estimations, which restricts the robustness of certain statistical tests. For instance, the kurtosis test in regression diagnostics is known to yield unreliable p-values with fewer than 20 observations. This limitation stems from data availability: while the ESG dataset from Refinitiv spans from 2002 to 2020, the associated news headline data used for sentiment estimation primarily covers the period from 2008 to 2024. As a result, only 13 full calendar years (2008–2020) were usable for combined ESG-sentiment modeling. Additionally, some tickers had shorter data spans due to missing values, as shown in Table 3.2.b, further reducing the effective sample.

There are several potential avenues to address this limitation in future research. The first and most straightforward solution is to extend the data collection period. By continuing to collect news data until at least the end of 2028, researchers can ensure 20 full years of overlap, meeting minimal statistical thresholds for time-series reliability.

Second, researchers could consider incorporating pre-2008 news data to align more fully with the ESG time series. Lee et al. (2024) demonstrate the viability of this approach using LexisNexis, which offers news coverage dating back to the 1980s [33][49]. By supplementing headline and article data from this archive, a more complete temporal alignment with Refinitiv's ESG dataset becomes possible. This would not only improve the model's statistical power but also support longer-term trend analysis.

Third, future research could move beyond annual ESG data by adopting methods for estimating daily ESG sentiment scores. While Refinitiv currently provides only annual ESG ratings, recent studies such as Dorfleitner and Zhang (2024) have proposed techniques for generating daily ESG proxies using BERT-based models based on access to both Refinitiv's ESG scores and an original dataset consisting of 245,723 raw news articles identified as ESG-related by Thomson Reuters [21]. Bert is one of the early pretrained large language models (LLM). By applying advanced language models (e.g., LoRA, LLaMA, GPT, PaLM, Falcon, Mistral), researchers can train to estimate daily ESG sentiment by reverse-engineering the relationship between ESG narratives in financial news (e.g., from Thomson Reuters Eikon) and official Refinitiv's ESG scores. This would significantly enhance the temporal granularity and responsiveness of ESG modeling. Furthermore, companies can generate ESG-friendly news titles and articles using Retrieval-Augmented Generation (RAG) or other types of LLMs to influence public perception and potentially enhance sentiment-linked market performance. This would significantly enhance the temporal granularity and responsiveness of ESG modeling.

Additionally, such methods directly address the trade-offs identified in Section 4.7 regarding dual sentiment aggregation strategies (_Davg and non-_Davg). By generating sentiment signals at the daily level, LLM-based ESG estimation can replicate or even improve upon the logic behind the _Davg approach, ensuring temporal consistency and reducing distortion from uneven news frequency. At the same time, these models retain the flexibility to capture cumulative sentiment exposure, similar to the non-_Davg approach, without relying on arbitrary aggregation windows. Thus, LLM-driven daily ESG estimation offers a scalable and data-driven solution that unifies the strengths of both aggregation strategies while minimizing their limitations.

## 7.2 Company-focused Limitations and Directions

An additional avenue for enhancing the robustness of future research is to increase the number of companies included in the analysis. The full Refinitiv ESG database contains data for approximately 16,000 companies, including around 14,500 public firms and 1,300 private entities, making it a rich resource for expanding sample size and representativeness [50].

Expanding the firm coverage offers two main benefits. First, it enables broader generalizability of the findings. With only 17 firms, the current sample size is limited and potentially biased by firm-specific effects. Incorporating a larger and more diverse set of firms—especially across different geographies and industries—would yield more statistically robust insights. Importantly, applying the same sentiment analysis and modeling methodology to a wider sample would primarily affect computational time, rather than requiring a methodological overhaul.

Second, a larger sample would support more nuanced and sector-specific interpretations of ESG–sentiment interactions. Firms can be grouped and compared by region, sector, industry, or size, offering the potential for richer subgroup analyses. For example, regional comparisons could illuminate differences in how ESG signals are interpreted across institutional or market contexts. Prior research by Naeem and Cankaya (2022) [51] found that the financial impact of ESG performance is more pronounced in developed markets than in emerging markets, particularly for firms operating in environmentally sensitive industries. Including a geographically diverse set of companies would allow future studies to test whether similar variations exist in ESG–sentiment interactions across different economic contexts.

Importantly, increasing the number of firms could also help address the relatively low incidence of triple significance models observed in this study, only 98 out of 1,360 possible ESG–sentiment combinations[1] (approximately 7.2%). While this reflects the study's rigorous statistical thresholds, it also suggests that expanding the dataset would likely increase both the number and diversity of significant models. A broader sample would improve model density and may reveal whether such interactions are more frequent in certain sectors, ESG dimensions, or regions. As a result, future research leveraging larger and more heterogeneous datasets could offer deeper insights into the conditional dynamics of sentiment and ESG on financial performance.

**Table 7.2. Sector and Industry Classification of Sampled Companies According to GICS (via Yahoo Finance)**

| Ticker | Company Name | Sector | Industry |
|---|---|---|---|
| AAPL | Apple Inc. | Technology | Consumer Electronics |
| AMZN | Amazon.com, Inc. | Consumer Cyclical | Internet Retail |
| BA | Boeing Company | Industrials | Aerospace & Defense |
| BAC | Bank of America Corp | Financial Services | Banks – Diversified |
| BPORD | BP p.l.c. | Energy | Oil & Gas Integrated |
| BNPQ | BNP Paribas SA | Financial Services | Banks – Regional |
| CVS | CVS Health Corp | Health Care | Healthcare Plans |
| F | Ford Motor Company | Consumer Cyclical | Automobile Manufacturers |
| GS | Goldman Sachs Group Inc. | Financial Services | Capital Markets |
| TOYOT | Toyota Motor Corp | Consumer Cyclical | Automobile Manufacturers |
| TENCE | Tencent Holdings Ltd | Communication Services | Internet Content & Information |
| NOKIA | Nokia Corp | Technology | Communications Equipment |

---

[1] (5 ESG scores)x(17 Companies from Refinitiv)x(16 Sentiment variable from Retrofitted Word Embeddings)

| PFE | Pfizer Inc. | Health Care | Drug Manufacturers – General |
| SALLI | Alibaba Group Holding Ltd | Consumer Cyclical | Internet Retail |
| SAMSU | Samsung Electronics Co., Ltd. | Technology | Consumer Electronics |
| TSLA | Tesla Inc. | Consumer Cyclical | Automobile Manufacturers |

One viable framework for sectoral classification is the Global Industry Classification Standard (GICS), which is widely adopted in both academic and professional finance contexts. Industry and sector data can be retrieved via Yahoo Finance, which applies the GICS system developed by MSCI (Morgan Stanley Capital International) and Standard & Poor's, providing a globally standardized taxonomy for firm classification (Royal, 2023) [52]. Table 7.2 demonstrates and presents the GICS-based sector and industry assignments of the 17 firms analyzed in this study, as sourced from Yahoo Finance.

The use of such standardized classifications also opens the door for testing industry-specific ESG sentiment patterns. For instance, results from Hypothesis 2 (H2) suggest that Samsung showed a particularly strong effect when its social score interacted with sentiment measures. This may indicate a country-specific sensitivity for Korean technology firms. In contrast, technology firms such as Apple and Nokia exhibited stronger significance when environmental scores were used, potentially reflecting different stakeholder expectations or regulatory environments. However, due to the limited number of companies per sector, these findings remain exploratory and cannot yet be generalized. A larger sample would allow for more rigorous testing of such patterns across industry clusters and national contexts.

## 7.3 Expanding Financial Variables Beyond Daily Returns

One notable limitation of this study is its reliance on a single financial indicator—daily adjusted closing price returns (Adj Close Adjusted close price adjusted for splits and dividend and/or capital gain distributions._R_DAvg)—as the primary dependent variable for modeling ESG and sentiment interactions. While short-term price fluctuations offer valuable insight into market reactions, they capture only a narrow aspect of firm performance and valuation. Future research could enhance this framework by incorporating broader financial indicators such as Tobin's Q, return on assets (ROA), return on equity (ROE), debt-to-equity ratios, or other metrics extracted from firms' 10-K filings.

This approach has been successfully implemented in previous studies. For example, Yu, J., Guo, Y., and Luu, B. V. (2018) [53] used Tobin's Q to evaluate how the informativeness of ESG disclosures—conditional on ESG ratings—affects firm valuation. Similarly, research by Xu et al. (2022) [54], Flammer (2013) [55], Intezar et al. (2024) [56], and Velte (2017) [57] examined other financial indicators like ROA and ROE as accounting-based indicators of ESG-financial performance relationships.

These financial variables allow for a richer interpretation of ESG's real economic impact—beyond market perception—by addressing firm fundamentals. Integrating such financial statement data would help bridge the gap between investor sentiment and operational outcomes, offering a more multidimensional and grounded evaluation of ESG's role in corporate performance.

# 8. Conclusion

This study provides robust evidence supporting Hypothesis 1 (H1), affirming that emotion-specific sentiment, when contextualized with ESG scores, significantly influences stock return behavior. Among 16 companies, 12 produced at least one triple-significant model, confirming the explanatory value of integrated ESG × Sentiment dynamics. Emotions like sadness, trust, and happiness emerged as particularly salient, frequently appearing in high-performing models ($R^2 > 0.6$), with Anticipation_Davg and Trust leading in prevalence across firms.

Hypothesis 2 (H2) also finds strong empirical backing. The moderating influence of sentiment on ESG–return relationships varies meaningfully across ESG subdimensions. Environmental and overall ESG scores consistently demonstrated the highest explanatory power, while economic, social, and governance categories showed more variable results. Sentiment modulation was not uniformly distributed across ESG categories or firms—some companies clustered heavily in specific ESG dimensions, pointing to potential industry-specific salience or investor expectations.

For Hypothesis 3 (H3), findings are more complex and mixed. Of the 197 statistically significant ESG × Sentiment interactions (after excluding neutral sentiments), only 94 (47.7%) aligned with the hypothesized direction, while 103 (52.3%) contradicted it. Moreover, the weighted average coefficient for the H3-consistent interactions was negative (–0.215; $p = 0.0005$), suggesting a net suppressive effect even among supportive cases. In contrast, the contradictory interactions showed a strong enhancing effect (average = 0.541; $p < 0.0001$). These results suggest that market responses to ESG-related sentiment are not symmetric: negative sentiment may exert a disproportionately strong dampening effect, while positive sentiment does not reliably amplify ESG-related financial outcomes.

Overall, the study confirms that integrating emotion-specific sentiment into ESG analysis provides a powerful and nuanced lens for understanding stock return behavior. The evidence suggests that Retrofitted Word Embeddings outperform traditional lexicons like NRC, particularly in terms of explanatory power and the clarity of sentiment signals. Furthermore, the dual sentiment aggregation strategy (_Davg and non-_Davg) adds methodological robustness by balancing temporal stability with information richness. These findings offer valuable insights for investors, policymakers, and researchers seeking to better understand the behavioral pathways through which ESG information is priced in financial markets.